\documentclass[aps,prx,reprint,superscriptaddress]{revtex4-2} 
\usepackage{graphicx,subfigure}
\usepackage{amsmath,amssymb}
\usepackage{mathtools}
\usepackage{multirow}
\usepackage{textcomp}
\usepackage{float}
\usepackage{xcolor}
\usepackage{jeffe}
\usepackage{ulem}
\usepackage{dsfont}
\usepackage{upgreek}
\usepackage{comment}
\usepackage{chngcntr}
\usepackage{hyperref}

\definecolor{bluegreen}{HTML}{16b5b2}

\newcommand{\avg}[1]{\langle #1 \rangle}

\newcommand{\ket}[1]{\ensuremath{\left\vert{#1}\right\rangle}}

\newcommand{\bra}[1]{\ensuremath{\left\langle{#1}\right\vert}}
\newcommand{\ketbra}[2]{\ensuremath{\left\vert{#1}\rangle\langle{#2}\right\vert}}
\newcommand{\braket}[2]{\ensuremath{\left\langle{#1}\vert{#2}\right\rangle}}

\newcommand{\uvec}[1]{\ensuremath{\hat{\mathbf{#1}}}}
\renewcommand{\abs}[1]{\ensuremath{\left\vert{#1}\right\vert}}

\definecolor{ao(english)}{rgb}{0.0, 0.5, 0.0}

\renewcommand{\vec}[1]{\ensuremath{\mathbf{#1}}}

\newcommand{\Id}{\ensuremath{\mathds{1}}}

\newcommand{\W}{W_1}
\newcommand{\Wtarget}{\ensuremath{W_*}}
\renewcommand{\th}{\ensuremath{\gamma}}
\newcommand{\sol}{\mathcal{A}}
\newcommand{\Nsol}{N_\sol}
\newcommand{\Topt}{T_\mathrm{opt}}
\newcommand{\Popt}{P_\mathrm{opt}}
\newcommand{\speedup}{Q}
\newcommand{\order}{O}

\newcommand{\Tmax}{T_\mathrm{max}}
\newcommand{\err}{\ensuremath{\varepsilon}}
\newcommand{\quantile}[2]{\left[#1\right]_{#2}}
\newcommand{\Jmax}{J_\mathrm{max}}
\newcommand{\Pthresh}{\mathcal{P}}
\newcommand{\Neff}{\Nsol^{\mathrm{eff}}}
\newcommand{\thopt}{\th_\mathrm{opt}}
\newcommand{\speedupopt}{Q_\mathrm{opt}}
\newcommand{\y}{\mu}
\newcommand{\rr}{r}
\newcommand{\wrms}{w_\mathrm{rms}}
\newcommand{\keff}{\ensuremath{k_\mathrm{eff}}}

\renewcommand{\R}{\ensuremath{\mathcal{R}}}
\newcommand{\Gmax}{\ensuremath{G_\mathrm{max}}}
\newcommand{\Ttotal}{\ensuremath{T_{\textrm{total}}}}
\newcommand{\z}{\ensuremath{\chi}}
\newcommand{\zbar}{\ensuremath{\overline{\chi}}}
\newcommand{\tmod}[1]{ \mathrm{mod} #1}
\newcommand{\Mod}[1]{\ \mathrm{mod} #1}

\begin{document}
\title{Number Partitioning with Grover's Algorithm in Central Spin Systems}

\author{Galit Anikeeva}
\thanks{These authors contributed equally to this work.}
\author{Ognjen Markovi\'{c}}
\thanks{These authors contributed equally to this work.}
\affiliation{Department of Physics, Stanford University, Stanford, California 94305, USA}
\author{Victoria Borish}
\thanks{These authors contributed equally to this work.}
\author{Jacob A. Hines}
\thanks{These authors contributed equally to this work.}
\affiliation{Department of Applied Physics, Stanford University, Stanford, California 94305, USA}
\author{Shankari V. Rajagopal}
\thanks{These authors contributed equally to this work.}
\author{Eric S. Cooper}
\author{Avikar Periwal}
\affiliation{Department of Physics, Stanford University, Stanford, California 94305, USA}
\author{Amir Safavi-Naeini}
\affiliation{Department of Applied Physics, Stanford University, Stanford, California 94305, USA}
\author{Emily J. Davis}
\author{Monika Schleier-Smith}
\affiliation{Department of Physics, Stanford University, Stanford, California 94305, USA}

\begin{abstract}
Numerous conceptually important quantum algorithms rely on a black-box device known as an oracle, which is typically difficult to construct without knowing the answer to the problem that the algorithm is intended to solve.  A notable example is Grover's search algorithm.  Here we propose a Grover search for solutions to a class of NP-complete decision problems known as subset sum problems, including the special case of number partitioning.  Each problem instance is encoded in the couplings of a set of qubits to a central spin or boson, which enables a realization of the oracle without knowledge of the solution.  The algorithm provides a quantum speedup across a known phase transition in the computational complexity of the partition problem, and we identify signatures of the phase transition in the simulated performance.  Whereas the naive implementation of our algorithm requires a spectral resolution that scales exponentially with system size for NP-complete problems, we also present a recursive algorithm that enables scalability. 
We propose and analyze implementation schemes with cold atoms, including Rydberg-atom and cavity-QED platforms.

\end{abstract}
\date{\today}

\maketitle

\section{Introduction}

Many quantum algorithms that offer a provable speedup over their best classical counterparts rely on the ability to query an oracle: a black box that knows the answer to the problem that the quantum computer is to solve.  A paradigmatic example is Grover's search algorithm~\cite{grover1996fast,grover1997quantum}, which theoretically speeds up the time to search through an unstructured database of $N$ entries, requiring only $\order(\sqrt{N})$ queries of the oracle rather than the classical $\order(N)$ queries.  By extension, Grover's algorithm can in principle speed up the search for solutions to a wide range of decision problems, including NP-complete problems~\cite{karp1972reducibility} such as boolean satisfiability, the clique problem, and the number partitioning problem~\cite{mertens1998phase, bernstein2013quantum}, with applications from cryptography to finance~\cite{merkle1978hiding,lyubashevsky2010public,weingartner1967methods,gilli2019numerical}.

Formally, any instance of a search or decision problem is represented by an oracle function $f(x)$ that acts on a string $x$ of $n$ bits and returns either 0 (failure) or 1 (success).  The search aims to find a value $X$ such that $f(X)=1$, while the decision problem asks whether such an $X$ exists at all.  In experimental demonstrations to date of Grover's search~\cite{chuang1998experimental,jones1998implementation,vandersypen2000implementation,kwiat2000grover,ahn2000information,anwar2004implementing,brickman2005implementation,walther2005experimental,prevedel2007high,barz2012demonstration,figgatt2017complete,godfrin2017operating,wu2019programmable}, implementing the oracle---a unitary operation controlled by $f(x)$---requires knowing the solution(s) $X$. To obtain a true benefit from a quantum algorithm involving an oracle, one requires a physical system that directly encodes the function $f$ in a manner that is agnostic to the solution~\cite{roget2020grover}.

\begin{figure}[th] 
    \centering
    \includegraphics[width=\columnwidth]{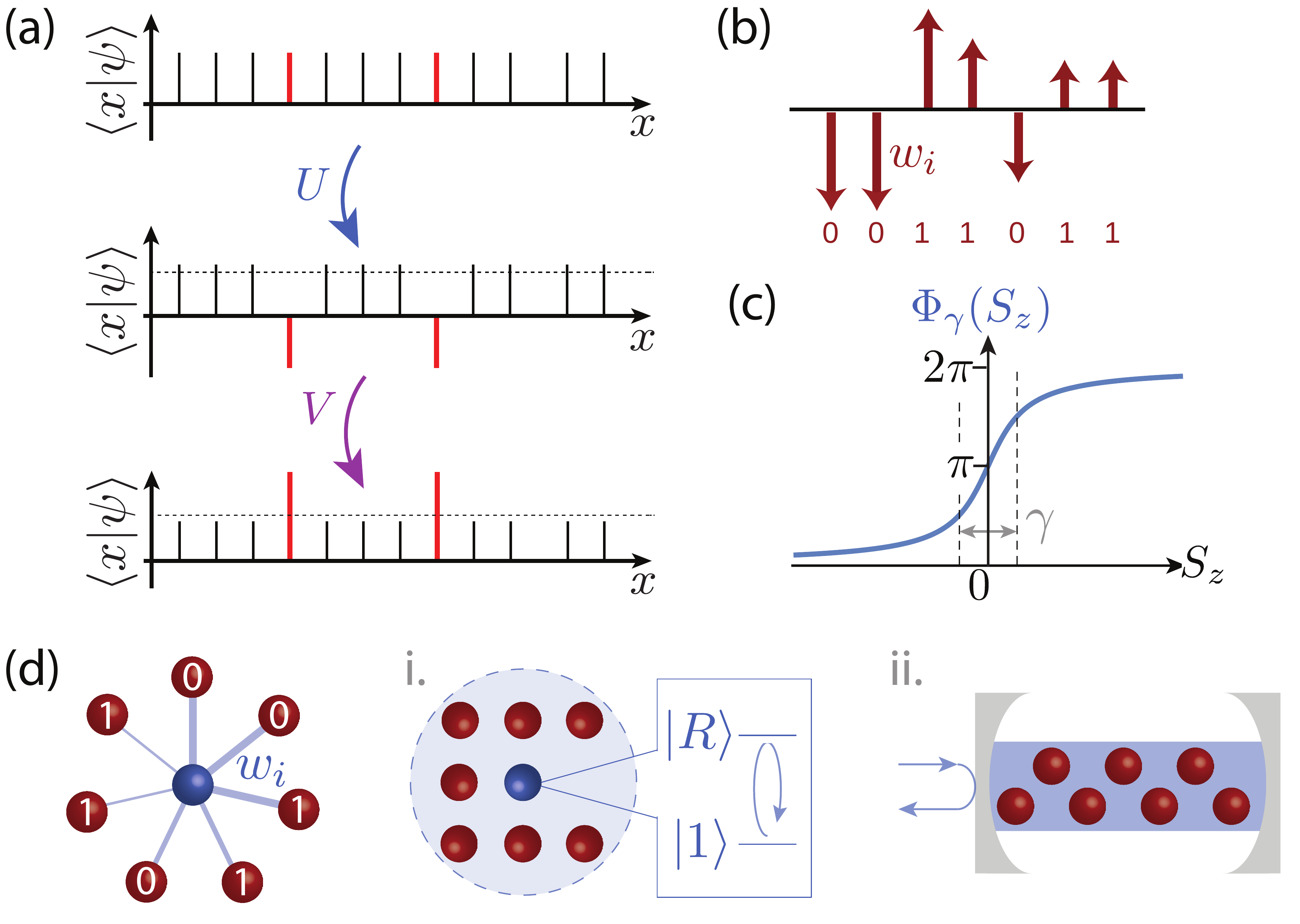} 
    \caption{(a) Sketch of Grover's algorithm, showing the amplitude of each basis state \ket{x} in the system state $\ket{\psi}$. One iteration consists of the oracle $U$ marking the solution states (red) with a $\pi$ phase shift, followed by inversion about the average $V$. (b) Number partitioning: a set of weighted spins is partitioned, if possible, into two sets of equal total weight.
    (c) Phase shift $\Phi_\th(S_z)$ applied by generalized oracle with step width $\th$. (d) The weights $w_i$ are encoded by couplings of system spins (red) to an ancilla (blue), which can be either (i) a central spin (e.g., Rydberg atom); or (ii) a bosonic mode (e.g., cavity).
    }
    \label{fig:setup}
\end{figure}

In this paper, we propose a genuine application of Grover's algorithm to solving the NP-complete number partitioning problem: \textit{Given n objects with integer weights, does there exist a bipartition that balances a scale?}
Our approach can be implemented in physical systems that take the form of either a central spin or central boson model, featuring $n$ qubits interacting with an ancilla spin or photon that plays the role of the oracle.  Crucially, the decision problem is encoded in the couplings of the qubits to the ancilla, allowing the oracle to be implemented without \textit{a priori} knowledge of the solution.  Numerical simulations of the quantum algorithm illustrate physical manifestations of a known phase transition in the computational complexity of number partitioning, including an exponential scaling of the spectral resolution required to solve hard problem instances.  A recursive variant of our algorithm avoids this exponential resource requirement, providing improved scalability.    By analyzing proposed implementations with Rydberg atoms and in cavity-QED systems, we show that a speedup is attainable in near-term experiments.

\section{Algorithm and Implementation}\label{sec:algoimplementation}

We begin with a brief review of Grover's algorithm [Fig.~\ref{fig:setup}(a)].  The algorithm starts by initializing a collection of $n = \log_2 N$ qubits in an equal superposition
\begin{equation}
\ket{\psi_0} = \frac{\left(\ket{0} + \ket{1} \right)^{\otimes{n}}}{2^{n/2}} = \sum_x c_{x,0} \ket{x}
\label{eq:psiInit}
\end{equation}
of all possible standard basis states labeled by $n$-bit numbers $x$, with $c_{x,0} = 1/\sqrt{N}$.  The objective is to amplify the amplitude $c_X$ of the solution state(s) $\ket{X}$.  To this end, the oracle $U$ first marks the solution(s) by applying a $\pi$ phase shift ($c_X \rightarrow e^{i\pi}c_X$) for all $X$ with $f(X)=1$.  The marked states are then amplified by inversion about the average: $c_x \rightarrow \overline{c} - (c_x - \overline{c})$ for all $x$, where $\overline{c} = \sum_x c_x / N$.  This inversion operation $V$ is accomplished by combining single-qubit Hadamard gates with an $n$-qubit controlled phase gate that is similar to the oracle but less technically demanding (see App.~\ref{sec:inversionOperator}), or can alternatively be replaced by single-qubit rotations only~\cite{jiang2017near}.  Thus, we focus on the challenge of realizing the oracle.

We will show a natural physical incarnation of the oracle for a class of decision problems known as subset sum problems~\cite{moore2011nature}, focusing on the special case of number partitioning.  We specify each problem instance by a list of $n$ weights $w_i\in (0,1]$ of finite bit depth $k$, and search for a partition into two sublists of equal total weight.  To encode the partition problem using $n$ qubits representing the objects with weights $w_i$, we let each qubit state indicate which subset ($\ket{0}$ or $\ket{1}$) an object is in [Fig.~\ref{fig:setup}(b)], so that the weighted collective spin 
\begin{equation}\label{eq:ss_cond_Sz}
S_z \equiv \frac{1}{2}\sum_i w_i \sigma^z_i
\end{equation}
represents the imbalance between the subsets.  Implementing the oracle then requires applying a $\pi$ phase shift to any $n$-qubit basis state $\ket{x}$ satisfying $S_z\ket{x} = 0$.

The quantum oracle thus requires implementing a collective phase gate $U = e^{i\pi f(x)} = e^{i\pi \delta(S_z)}$, where $\delta(\cdot)$ denotes the Kronecker delta function.  To design a physical implementation of this gate, it is helpful to define a generalized oracle $U_\th = e^{i\Phi_\th(S_z)}$ in terms of an $S_z$-dependent phase shift
\begin{equation}\label{eq:Phi}
\Phi_\th (S_z) = 2\arctan\left(2 S_z/\th\right) + \pi,
\end{equation}
which steps from zero to $2\pi$ as a function of $S_z$ and provides a $\pi$ phase shift at $S_z = 0$ [Fig.~\ref{fig:setup}(c)].  The ideal oracle is obtained in the limiting case $U \equiv U_{\th\rightarrow 0}$ of an infinitely steep phase step.

The collective phase gate $U_\th$ can be enabled by coupling the qubits to an ancilla, which may take the form of an auxiliary qubit or a bosonic mode.  We consider either a central spin model
\begin{equation}\label{eq:central_spin}
H_q = J_\mathrm{max} I_z S_z
\end{equation}
featuring an ancilla qubit represented by a spin-1/2 operator $I_z$, or a central boson model
\begin{equation}\label{eq:central_boson}
H_c = J_\mathrm{max} c^\dagger c S_z
\end{equation}
featuring a cavity mode with annihilation operator $c$.  In both cases, the ancilla couples to $n$ system spins in the starlike graph of Fig.~\ref{fig:setup}(d), and hence to the weighted collective spin $S_z$. The maximum coupling between a system spin and the ancilla is parameterized by $\Jmax$.

For concreteness, we describe  representative implementations of the central boson and central spin models with cold atoms [Fig.~\ref{fig:setup}(d)].  The system spins are encoded in two internal states $\ket{0}$, $\ket{1}$ and coupled to either a cavity mode~\cite{jiang2008anyonic,chen2015carving,davis2018painting,gleyzes2007quantum,welte2018photon,mcconnell2015entanglement,barontini2015deterministic,davis2020protecting} or an auxiliary atom that can be excited to a Rydberg state~\cite{saffman2009efficient,molmer2011efficient,zhang2010deterministic,wilk2010entanglement,jau2016entangling,zeiher2016many,zeiher2017coherent,picken2018entanglement,borish2020transverse,madjarov2020high,levine2019parallel,young2020asymmetric,ashida2019quantum}. Each coupling $w_i \Jmax$ represents the energy shift of the $\ket{0}\rightarrow\ket{1}$ transition in atom $i$ when either a photon enters the cavity or the auxiliary atom is excited.  In the cavity implementation, the photon imparts an ac Stark shift~\cite{jiang2008anyonic,chen2015carving,davis2018painting,gleyzes2007quantum,mcconnell2015entanglement,welte2018photon}.  In the Rydberg implementation, the excited ancilla suppresses an ac Stark shift induced by classical control fields that couple the system atoms' state $\ket{1}$ to a Rydberg state.  In both cases, the weights $w_i$ can be programmed via the atomic positions or control fields. The net effect of the couplings $w_i \Jmax$ on the ancilla is a frequency shift $\Jmax S_z$ that depends on the weighted collective spin $S_z$.

The $S_z$-dependent resonant frequency of the ancilla is crucial to enabling the oracle.  In the central boson model, the oracle relies on the phase response of a driven harmonic oscillator.  For a one-sided cavity of linewidth $\kappa$, the output field is phase shifted by $\pi$ for a resonant drive compared with the off-resonant case.  Having the drive field consist of a single photon that is resonant if and only if $S_z=0$ yields precisely the oracle operation $U_\th$, with a phase step of dimensionless width $\th = \kappa / J_\mathrm{max}$, where we set $\hbar=1$.  In the central spin model, the oracle $U_\gamma$ is implemented by attempting to drive a $2\pi$ rotation of the ancilla with a field that is resonant if the weighted spin $S_z$ is zero.  For a suitably shaped drive pulse, the ancilla atom ends up in its initial state irrespective of $S_z$~\cite{rosen1932double}, and the entire system acquires a $\pi$ geometric phase shift only when $S_z=0$.  The width $\th = \kappa / J_\mathrm{max}$ of the phase step is now set by the bandwidth $\kappa=2\pi/\tau$ of the pulse with temporal width $\tau$.

\begin{figure}
	\centering
	\includegraphics[width=\columnwidth]{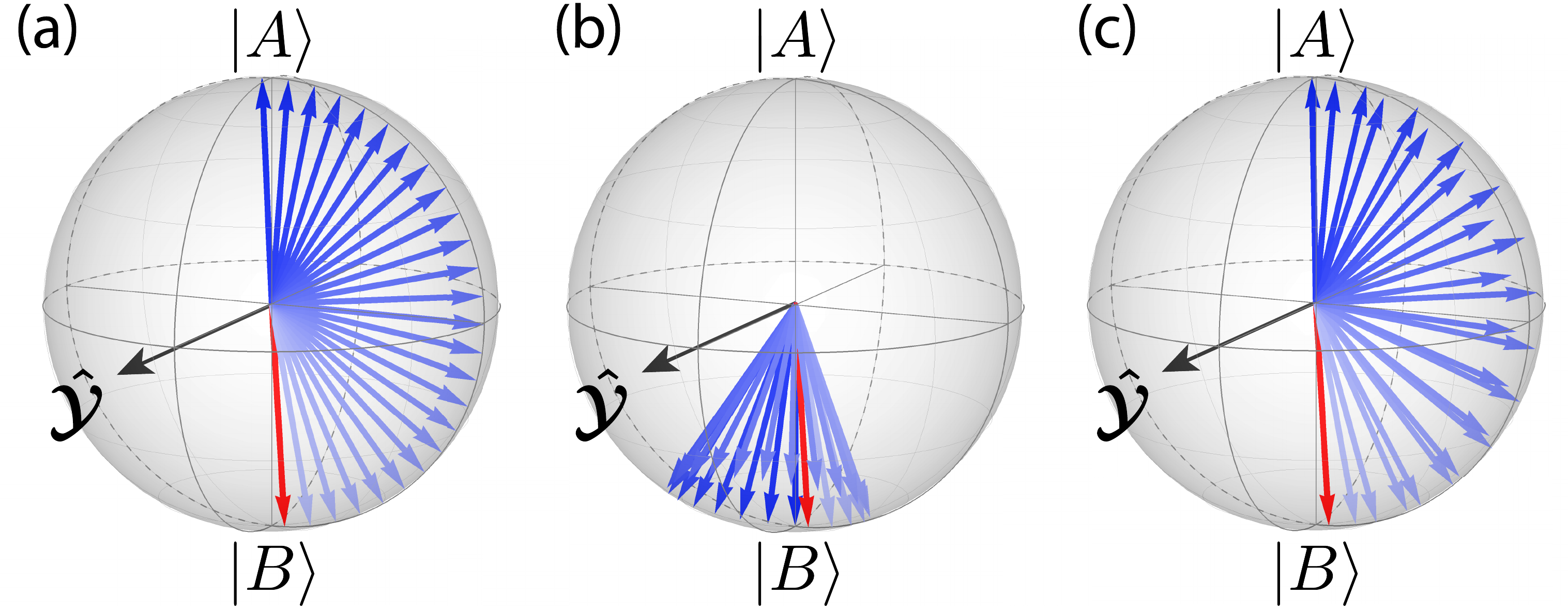}
	\caption{Visualization of Grover's algorithm and generalized oracle.  (a) Grover's algorithm with ideal oracle for $N = 2^{10}$ and $\Nsol = 1$.  Over repeated iterations (blue), the state $\ket{\psi_0}$ (red) approaches the solution state $\ket{A}$. (b) Grover's algorithm with naive application of the generalized oracle for $\epsilon = 0.25$. (c) The spin-echo sequence compensates for the imperfection of the oracle, allowing similar performance to the ideal case.
	\label{fig:groverBlochSpheres}}
    \end{figure}

To examine the performance of the generalized oracle, we first introduce a convenient visualization of Grover's algorithm~\cite{nielsen2010quantum}.  We define the solution space $\sol = \{\ket{X} : S_z\ket{X} = 0\}$
as the set of states that solve the partition problem and let
\begin{equation}
\ket{A} = \frac{1}{\sqrt{\Nsol}}\sum_{\ket{X}\in \sol} \ket{X}
\end{equation}
denote the equal superposition of all solutions (assuming their existence) where $\Nsol$ is the number of solutions.  We additionally define an orthogonal state
\begin{equation}
\ket{B} \propto \ket{\psi_0} - \ket{A}\braket{A}{\psi_0},
\end{equation}
where $\ket{\psi_0}$ is the initial state of Eq.~\ref{eq:psiInit}.  The states $\ket{A}$ and $\ket{B}$ span an SU(2) subspace that can be visualized on a Bloch sphere with $\ket{A}$ and $\ket{B}$ as poles.

Grover's algorithm ideally takes place entirely within this subspace of the full $2^n$-dimensional Hilbert space, iteratively rotating the initial state $\ket{\psi_0}$ towards the solution state $\ket{A}$.  Each iteration
\begin{equation}
\ket{\psi_{T+1}} = VU\ket{\psi_T},
\end{equation}
comprises the oracle $U$ and inversion about the average $V$.  The net effect of these two operations is a rotation about the $\boldsymbol{\hat{\mathcal{Y}}}$ axis [Fig.~\ref{fig:groverBlochSpheres}(a)]. For $\Nsol/N\ll 1$, a near-unity success probability is achieved after an optimal number of iterations
\begin{equation}\label{eq:Tgrover}
T \approx (\pi/4) \sqrt{N/\Nsol}.
\end{equation}

The generalized oracle with a nonzero step width introduces an error that, to lowest order, is correctable by spin echo. To visualize how, we consider a simplified scenario where there exist only two possible values of the phase $\Phi_\th \in \{\epsilon,\pi\}$. For nonzero $\epsilon$, the combination of the generalized oracle and inversion about the average induces the state to rotate about a tilted axis [Fig.~\ref{fig:groverBlochSpheres}(b)].  To mitigate accumulation of error, we alternate between applying the oracle $U_\th$ and its Hermitian conjugate $U_\th^\dagger$.  A pair of two Grover iterations then takes the form
\begin{equation}\label{eq:hobbling}
\ket{\psi_{T+2}} = VU_\th^\dagger V U_\th\ket{\psi_T}
\end{equation}
where $U_\th^\dagger = \R^\dagger_{\pi}(\uvec{x}) U_\th \R_{\pi}(\uvec{x})$ is implemented by a spin-echo sequence involving two global $\pi$ rotations $\R_{\pi}(\uvec{x})$ about the individual qubits' $\uvec{x}$ axes.  The result is the trajectory shown in Fig.~\ref{fig:groverBlochSpheres}(c), which achieves similar performance to the ideal oracle in Fig.~\ref{fig:groverBlochSpheres}(a).

Even with spin echo, the step width will ultimately limit the resolution of the generalized oracle: selectively amplifying only spin configurations with $S_z = 0$ requires a narrow step.  Further, producing a narrow step requires a long coherence time, so that dissipation will place physical limits on the performance of the algorithm.  We elaborate on both of these considerations in Secs.~\ref{sec:speedup} and~\ref{sec:dissipation}.  First, however, we examine the application of Grover's algorithm to number partitioning using a phase step narrow enough to resolve even the least significant bit of the weights.

\section{Speedup in Number Partitioning}\label{sec:speedup}

\begin{figure}[t]
    \centering
    \includegraphics[width=\columnwidth]{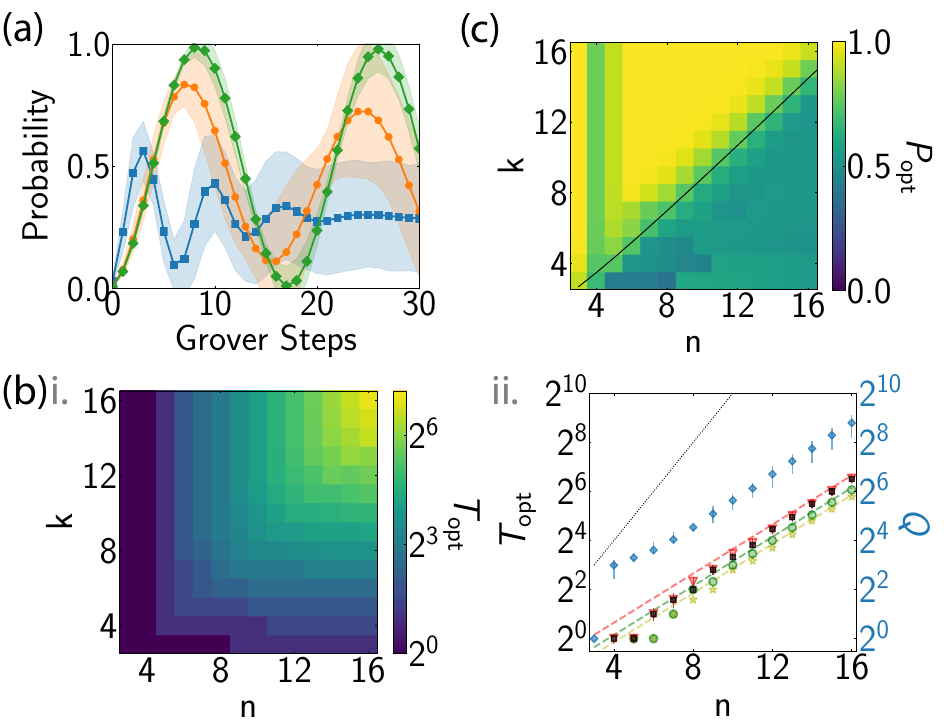} 
    \caption{Number partitioning with generalized oracle of step width $\th = 2^{-k}$. (a) Success probability $P_T$ for $n=8$, $k=4,8,12$ (blue squares, orange circles, and green diamonds). Shading indicates standard deviation over 5000 instances of the weights. (b.i) Optimal number of iterations $\Topt$ versus $(n,k)$. (b.ii)  $\Topt$ for $n=k$, grouped by number of solutions $N_A = 2, 4, 6$ (red triangles, green circles, and yellow stars) and compared with asymptotic theory $\Topt \propto \sqrt{N}$ (dashed lines).  Black squares show average over all instances. Dotted gray line indicates linear scaling $\Topt \propto N$. Blue diamonds show median speedup $\quantile{Q}{0.5}$ for $n=k$, with error bars indicating interquartile range. (c) Probability $\Popt$ versus $(n,k)$. Black line shows critical bit depth $k_c(n)$.}
    \label{fig:partition_results}
\end{figure}

To analyze the performance for number partitioning, we generate sets of $n$ random $k$-bit weights and postselect for instances where at least one perfect partition exists.  For each such instance, we calculate the success probability
\begin{equation}
P_T = \sum_{\ket{X}\in \sol} \abs{\braket{X}{\psi_T}}^2
\end{equation}
as a function of the number $T$ of calls to the oracle, applied with spin echo (Eq.~\ref{eq:hobbling}).  Figure~\ref{fig:partition_results}(a) shows examples of $P_T$ for $n = 8$ spins, bit depths $k = 4, 8, 12$, and a step width $\gamma = 2^{-k}$ just narrow enough to resolve changes in the least significant bit of $S_z$.  As expected from the Bloch-sphere picture, the success probability oscillates as a function of $T$.  The maximum probability and the time to reach it combine to determine the effectiveness of the algorithm.

As a single figure of merit, we calculate the total number of calls to the oracle required to reach a specified (near-unity) success probability $\Pthresh$. For a search procedure with fixed success probability $P$ per trial, the number of trials $M$ needed to reach a probability $\Pthresh = 1-\varepsilon$ of finding a solution is
\begin{equation} \label{eq:bernoulli}
    M(P, \varepsilon) = \frac{\ln{(\varepsilon)}}{\ln{(1-P)}}.
\end{equation}
Thus, reaching the target error $\err$ with Grover's algorithm requires querying the oracle a total of $\Ttotal=M(P_T, \varepsilon)T$ times. To minimize this quantity, we first calculate its median value as a function of $T$ over many instances of weights at a given $(n,k,\th)$. We then define $\Topt$ as the number of Grover iterations that minimizes the median total number of queries $\Ttotal$. Note that $\Topt$ is independent of the target error $\varepsilon$.

Figure~\ref{fig:partition_results}(b.i) shows the optimal number of queries $\Topt$ as a function of the number of spins $n$ and bit depth $k$, at fixed step width $\gamma=2^{-k}$.  The scaling of $\Topt$ with $n$ is shown in Fig.~\ref{fig:partition_results}(b.ii) for a cut at $n=k$ (black squares), where the number of perfect partitions is typically of order one~\cite{mertens1998phase}.  We additionally plot $\Topt$ for instances of the weights postselected according to the number of solutions $\Nsol = 2, 4, 6$ (red triangles, green circles, and yellow stars).  In each case, the optimal number of iterations approaches the prediction of Eq.~\ref{eq:Tgrover} (dashed lines) at large $N = 2^n$, scaling as $\Topt\propto \sqrt{N}$. Quantifying the resulting speedup requires additionally examining $\Popt$, the success probability after $\Topt$ iterations [Fig.~\ref{fig:partition_results}(c)].

The dependence of success probability $\Popt$ on $(n,k)$ reflects a known phase transition in the computational complexity of the number partitioning problem~\cite{mertens1998phase,Borgs01phasetransition}.  For small bit depth $k\lesssim n$ (the ``easy'' phase), there typically exist many perfect partitions.  For large bit depth $k\gtrsim n$ (the ``hard'' phase), perfect partitions are rare and thus---even when postselecting for their existence---the probability of finding them by random guessing is exponentially small in $n$.  By contrast, in our quantum search [Fig.~\ref{fig:partition_results}(c)], the success probability $\Popt$ is everywhere of order unity and highest in the ``hard'' phase, since Grover's algorithm is most effective when solutions are few.  The phase boundary lies at a critical bit depth~\cite{mertens1998phase} 
\begin{equation}\label{eq:kc}
    k_c(n) \equiv n -\frac{1}{2} \log_2\left(\frac{n\pi}{6}\right),
\end{equation}  
shown by the black curve in Fig.~\ref{fig:partition_results}(c), where the average number of perfect partitions is $\avg{\Nsol} \sim \sqrt{6/(\pi n)}2^{n-k} = 1$~\cite{Mezard2009}.

We quantify the advantage of the algorithm by calculating the limited quantum speedup $\speedup$, defined as in Ref.~\cite{ronnow2014defining} by comparing the quantum search with an algorithmically similar classical search. The most analogous classical algorithm is a memoryless search, which at each trial samples (with replacement) a random partition with success probability $P_0=\Nsol/N$. The number of memoryless search trials needed to reach a target success probability $\Pthresh$ also follows from Eq.~\ref{eq:bernoulli}. For each problem instance, we define speedup $\speedup$ as the ratio of memoryless trials to total Grover iterations: 
\begin{equation}\label{eq:speedup_scaling}
    Q = \frac{1}{\Topt}\frac{\ln{(1-\Popt)}}{\ln{(1-\Nsol/N})}.
\end{equation}
This speedup is independent of the target error $\varepsilon$, thanks to the algorithmic similarity of the two memoryless search algorithms, as further discussed in App.~\ref{sec:classicalAlgorithms}. Figure~\ref{fig:partition_results}(b.ii) shows the median speedup $\quantile{Q}{0.5}$, where $\quantile{Q}{q}$ denotes the $q$th quantile over problem instances. We observe the expected scaling $Q\propto \sqrt{N}$ of the speedup in query complexity.

\begin{figure*}[t]
    \includegraphics[width=\textwidth]{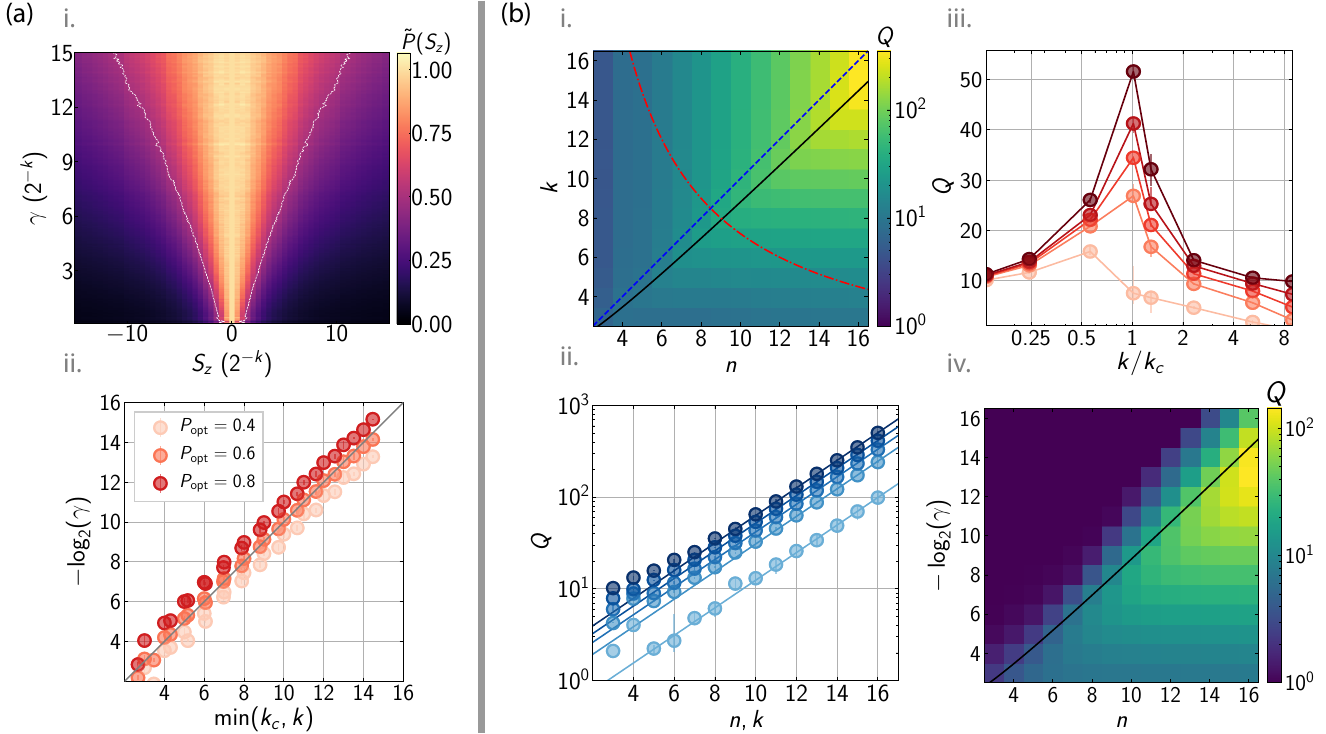}
    \caption{(a) Step width as capture range.
    (a.i) Normalized probability distribution $\tilde{P}(S_z)$ versus $\th$, for $n = k = 6$ with no postselection. White lines indicate contours of $\tilde{P}(S_z)$=0.5.  
    (a.ii) Step width $\th$ required to obtain $\Popt = [0.4,0.6,0.8]$, shaded from lighter to darker.  Results are plotted versus $\min(k_c,k)$, with markers showing average over all $(n,k)$ with $3 \leq n,k\leq 16$.
    Gray line shows $\th_c = 2^{-\min(k_c,k)}$.  
    (b) Quantum speedup in the decision and optimization problems.
    (b.i) Median speedup $\quantile{Q}{0.5}$ versus $(n,k)$ for the decision problem at step width $\th_c$. Lines denote $k=k_c$ (solid black),  $k=n$ (dotted blue), and fixed problem size $nk = 72$ (dashed red).
    (b.ii) Cuts of $Q$ along $n=k$ for different quantiles $q = [0.01, 0.25, 0.5, 0.75, 0.99]$, shaded from lightest to darkest. Lines denote $\sqrt{N}$ scaling.
    (b.iii) Cuts of $Q$ at $nk=72$ [red dashed line in (b.i)] for different quantiles, as in (b.ii). Lines are a guide to the eye.
    (b.iv) Median speedup $\quantile{Q}{0.5}$ versus $n$ and $\keff = -\log_2{\gamma}$ for the optimization problem with machine-precision weights, approximating the large-$k$ limit. Black line shows $-\log_2\th = k_c$.}
    \label{fig:speedup_results}
\end{figure*}

A caveat is that physical limitations might preclude successfully implementing the algorithm in cases requiring a narrow step width $\gamma$.  We have so far assumed a step width $\gamma = 2^{-k}$, motivated by the intuition that $\gamma$ sets a capture range of $S_z$ values amplified by Grover's algorithm.  To test this intuition, we plot the normalized probability distribution $\tilde{P}(S_z) \equiv P(S_z)/P(S_z=0)$ after $\Topt(\th)$ Grover iterations as a function of step width $\gamma$ [Fig.~\ref{fig:speedup_results}(a.i)], for $n=k=6$ without postselecting on the existence of perfect partitions.  Consistent with our expectation, the width of the distribution is approximately set by the step width $\gamma$.  An analytic derivation of this capture range is given in App.~\ref{sec:captureRange}.

To capture only true solutions $S_z=0$, the step width $\gamma$ should be smaller than the smallest nonzero $\abs{S_z}$ value.  In the easy regime $k\lesssim n$, a step width $\th\lesssim 2^{-k}$ is required to distinguish $S_z=0$ from $S_z=\pm 2^{-k}$.  However, with increasing $k$, the typical size of the smallest residue approaches a finite value $\abs{S_z} \approx 2^{-k_c}$~\cite{mertens1998phase}.  Thus, the critical bit depth $k_c(n)$ in Eq.~\ref{eq:kc} represents the resolution required to discriminate the smallest typical residue $\abs{S_z}$ in the large-$k$ limit.  For arbitrary $(n,k)$, we can choose the oracle to have resolution
\begin{equation}\label{eq:gammac}
-\log_2\th_c = \min(k_c,k) \approx \min(n,k),
\end{equation}
coarser than we have so far assumed in the hard regime. We verify Eq.~\ref{eq:gammac} by plotting the resolution $-\log_2\th$ required to reach a fixed success probability $\Popt$, averaging over all pairs $(n,k)$ with $3\leq n,k \leq 16$, in Fig.~\ref{fig:speedup_results}(a.ii).  For each of three different values of $\Popt = 0.4, 0.6, 0.8$, the required step width $\gamma$ is within a constant factor of $\gamma_c$.

We plot the quantum speedup for this less stringent choice of step width $\th_c$ in Fig.~\ref{fig:speedup_results}(b.i).  The speedup exhibits a maximum along the phase boundary $k=k_c(n)$ (solid black curve).  In Fig.~\ref{fig:speedup_results}(b.ii), we examine the scaling of the speedup along an approximation to this curve chosen to ensure integer values of $(n,k)$, namely, the $n=k$ cut (dotted blue line).  We plot the speedup $\quantile{Q}{q}$ versus $N$ for different quantiles $q$ (blue circles) and find good agreement with an asymptotic scaling $Q\propto \sqrt{N}$ (solid lines) for all quantiles.  Thus, the generalized oracle with the critical step width $\gamma_c$ suffices to achieve an $\order{(\sqrt{N}})$ speedup, the same scaling that is achieved by the ideal oracle and proven to be optimal for an unstructured search~\cite{bennett1997strengths,boyer1998tight,zalka1999grover}.

The phase transition in computational complexity manifests in a sharp peak in the speedup at the phase boundary $k = k_c(n)$.  We observe this peak in Fig.~\ref{fig:speedup_results}(b.iii) along a cut of fixed problem size $nk$, i.e., fixing the total number of bits encoding the set of $n$ weights. The peak in the speedup reflects the known result that the hardest problem instances are not deep in the hard regime, but rather near the phase transition \cite{mertens1998phase, impagliazzo1996efficient}. In particular, the hardest problems are those with the largest ratio $N/\Nsol$ of the size of the search space to the number of solutions, after postselecting for the existence for solutions. This ratio reaches a maximum near the phase boundary, explaining the peak in $\speedup \propto \sqrt{N/\Nsol}$.

Even in the experimentally relevant case where the weights are not restricted to a finite bit depth, the resolution of the oracle sets an effective bit depth $\keff = -\log_2\gamma$ that can reveal the complexity phase transition.  For real-numbered weights $w_i \in (0,1]$, we consider the optimization problem of minimizing $\abs{S_z}$, defining the success probability $P_\mathrm{opt}$ as that of finding the system in a configuration of minimal $\abs{S_z}$ after an optimal number of Grover iterations.  We plot the median speedup $\quantile{Q(n,\gamma)}{0.5}$ in Fig.~\ref{fig:speedup_results}(b.iv). As a function of $\keff$ at fixed $n$, the speedup first rises to a maximum at $\keff \approx k_c$ before declining precipitously for $\keff > k_c$ due to the narrowness of the capture range, providing a striking signature of the complexity phase transition.

\section{Effects of Dissipation}\label{sec:dissipation}

A key challenge for experimental implementations is that producing a narrow phase step requires a long coherence time. 
Specifically, at fixed interaction strength $\Jmax$, the step width $\th$ determines the physical time $\kappa^{-1}\sim 1/(\th\Jmax)$ to implement the oracle operation $U_\th$.  Even a single error occurring during this time thwarts the amplification process.  For concreteness, we consider an error model in which the excited ancilla decays---or, equivalently, the ancilla photon is lost---at rate $\Gamma_a$.  In terms of the interaction-to-decay ratio $\rho\equiv \Jmax/\Gamma_a$, the error rate per query of the oracle is then approximately $\Gamma_a/\kappa = 1/(\rho\gamma)$.  Thus, on average $\Tmax \sim \rho \th$ Grover iterations can be implemented before incurring an error.  For $\rho\gamma_c \lesssim \Topt$, the algorithm must be run at an increased step width $\th>\th_c$ that reduces the speedup.

\begin{figure}[t]
    \centering
    \includegraphics[width=\columnwidth]{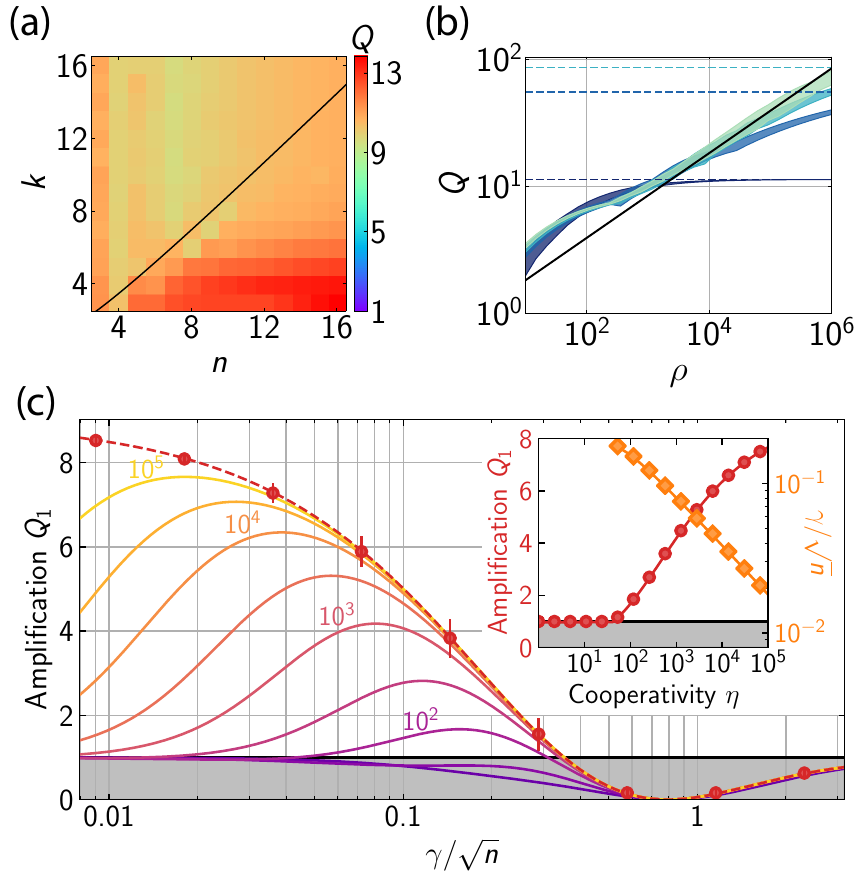}
    \caption{Effects of dissipation.
    (a) Median speedup $\quantile{Q}{0.5}$ versus $(n,k)$ in the presence of decay with interaction-to-decay ratio $\rho=10^3$. Solid black line denotes $k=k_c$.
    (b) Median speedup $\quantile{Q}{0.5}$ versus $\rho$ for $n=k=(4,6,8,10)$ denoted by dark blue to light green shaded lines. The shading denotes the interquartile range. Black solid line denotes scaling $Q \sim \rho^{1/3}$. Dashed lines show maximum achievable $Q$ for each system size $n$.  (c) Amplification versus step width $\gamma$ for $T=1$.  Solid curves show average amplification in large-$N$ limit for finite cavity cooperativity $\eta = 10^1,3\times 10^1,10^2, 3\times 10^2, \dots, 10^5$ (purple to yellow) and for unitary evolution (red dashed).
    Dark red circles show simulated amplification at $n=k=12$ with no dissipation; error bars denote standard deviation. Inset shows optimal amplification (red circles) and step width (orange diamonds) versus $\eta$ for $n=k=12$, matching the prediction for large $N$ (solid curves).}
    \label{fig:decay}
\end{figure}

Figure~\ref{fig:decay}(a) shows the speedup calculated at finite interaction-to-decay ratio $\rho=10^3$.  We model the decay by modifying the frequency shift of the ancilla's excited state (Sec.~\ref{sec:algoimplementation}) with an imaginary component, $\Jmax S_z + i\Gamma_a/2$, as detailed in App.~\ref{app:dissipation}.  We choose the step width $\th$ for each $(n,k)$ to maximize the speedup, accounting for a reduction in success probability due to the chance of ancilla decay. While the speedup no longer achieves $\order{(\sqrt{N})}$ scaling, we preserve an advantage $Q\approx 10$ compared with the classical search.  The dependence of the speedup on interaction-to-decay ratio $\rho$ is shown in Fig.~\ref{fig:decay}(b) for $n=k$ at different system sizes $n$.  The speedup scales as $Q\sim \rho^{1/3}$, consistent with an analytic model derived in App.~\ref{app:dissipation}, before saturating to the value expected for the ideal Grover's algorithm.

An interaction-to-decay ratio $\rho\gtrsim 10^3$ is experimentally accessible in an implementation of the central spin model using Rydberg atoms, as detailed in App.~\ref{sec:centralSpinModel}.  In this implementation, the dominant dissipative process is decay of the ancilla from the Rydberg state, whereas decay of the system spins is suppressed by coupling to their Rydberg states off-resonantly~\cite{zeiher2016many,zeiher2017coherent,jau2016entangling,borish2020transverse}. In terms of the maximum attainable Rabi frequency $\Omega_\mathrm{max}$ of this coupling, the interaction-to-decay ratio is limited to $\rho < \Omega_\mathrm{max}/(2\sqrt{n} \Gamma)$, which permits values of order $\rho\sim 10^3$ for realistic laser powers and high-lying Rydberg states.

At lower interaction-to-decay ratios, the optimum speedup is obtained by performing only a single Grover iteration.  In the absence of dissipation, this single-cycle speedup $Q_1$ is identical to the amplification factor $P_1/P_0$, assuming $P_{0,1} \ll 1$.  Figure~\ref{fig:decay}(c) shows $Q_1=P_1/P_0$ as a function of step width $\gamma$ for $n=k=12$ with no dissipation (red circles), corroborating an analytical model derived in App.~\ref{sec:captureRange} in the large-$N$ limit (dashed curve).  The model shows that the gain is set by $\gamma/\sqrt{n}$, which parameterizes the ratio of the step width to the width of the initial $S_z$ distribution, and saturates at a maximum value $Q_1=9$ for $\gamma/\sqrt{n} \ll 1$. Ancilla decay reduces the amplification $Q_1$ below this ideal curve, becoming significant for interaction-to-decay ratios $\rho \lesssim 1/\gamma$.  The optimum speedups in Fig.~\ref{fig:decay}(b) are obtained from a single amplification cycle for interaction-to-decay ratios $\rho\lesssim 10^2$.

A single amplification cycle could be performed in near-term realizations of the central boson model with atoms in a cavity (App.~\ref{sec:centralBosonModel}), by driving with a weak coherent field and heralding on the detection of a photon.  The coherence of the atom-cavity coupling is quantified by the cooperativity $\eta = 4g^2/\kappa\Gamma_e$, where $g$ is the vacuum Rabi frequency and $(\kappa,\Gamma_e)$ are the linewidths of the cavity and an atomic excited state to which it couples.  The resulting interaction-to-decay ratio scales as $\rho \propto \eta\gamma/n$, reflecting the fact that decreasing the dimensionless step width $\gamma = \kappa/\Jmax$ comes at the cost of increasing the photon loss probability by atomic scattering.  Achieving amplification requires reaching a step width $\gamma < \sqrt{n/12}$ narrower than the initial $S_z$ distribution while keeping $\rho\gamma > 1$ to avoid photon absorption, and hence requires strong coupling $\eta \gg 1$.

The full dependence of amplification $Q_1$ on step width $\gamma$ and cooperativity $\eta$ is shown by the solid curves in Fig.~\ref{fig:decay}(c).  Notably, the amplification at an optimal step width [Fig.~\ref{fig:decay}(c) inset] 
is independent of the number of spins $n$, depending only on the cooperativity $\eta$.  A state-of-the-art optical cavity with demonstrated cooperativity $\eta \sim 200$~\cite{colombe2007strong} thus allows for amplifying solutions to the partition problem at scalable system size.  Stronger amplification could be attained by coupling Rydberg atoms or superatoms~\cite{zeiher2015microscopic,paris2017free} to a high-cooperativity millimeter-wave cavity~\cite{gleyzes2007quantum,haroche2006exploring,suleymanzade2020tunable}.  For the parameters of Ref.~\cite{gleyzes2007quantum}, the cooperativity $\eta = 4\times 10^8$ is no longer the limiting factor.  Instead, finite lifetime $\Gamma^{-1}$ of the Rydberg states places a limit $\rho < g/(n^{3/2}\Gamma) = 5\times 10^3 / n^{3/2}$ on the interaction-to-decay ratio, which permits near-maximal $Q_1$ for up to $n\sim 30$ atoms.  Rydberg-based implementations might be further enhanced by inhibition of spontaneous emission~\cite{hulet1985inhibited,nguyen2018towards}.

\section{Scalable Algorithm}\label{sec:recursive}

The requirement of an exponentially fine resolution of the oracle poses challenges for scalability in the simple application of Grover’s algorithm presented so far.  Specifically, we showed in Sec.~\ref{sec:speedup} that the required step width $\th\sim 2^{-k}$ becomes exponentially small with increasing system size $n\sim k$ for the hardest problem instances.  As a result, if we scale the system-ancilla couplings such that the energy grows extensively with system size by fixing the maximum coupling $\Jmax$, then the time required for each query of the oracle grows as $2^k\sim 2^n$.  Alternatively, to keep the query time fixed, the energy must be chosen to grow exponentially with increasing system size.

\begin{figure}[t]
    \centering
    \includegraphics[width=\columnwidth]{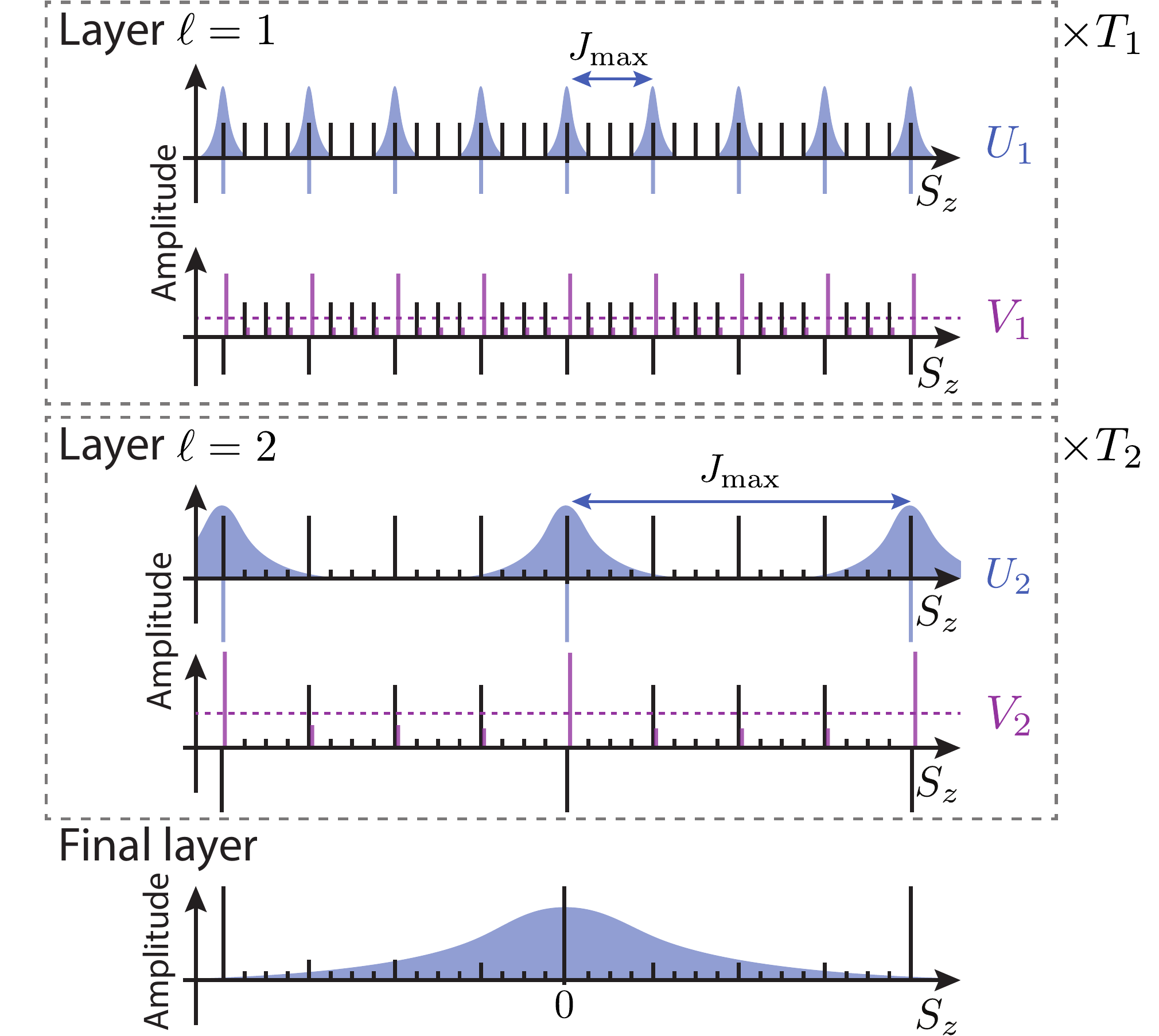}
    \caption{Sketch of the scalable algorithm, showing amplitudes of basis states versus $S_z$. Each layer $\ell$ of the algorithm consists of $T_\ell$ amplification cycles, each comprising the modular oracle $U_{\ell}$ and recursive inversion operator $V_{\ell}$. The modular oracle acts on states spaced in energy by $\Jmax$ with a spectral resolution $\gamma\Jmax$ illustrated by the shaded blue curves. The operator $V_\ell$ inverts the amplitudes of the states amplified by the previous layer of the algorithm about their average (dashed purple line). Implementing $V_\ell$ requires recursion to the lower layers of the algorithm. In the final layer $\ell = k/m$, the standard nonmodular oracle is used.
    }
    \label{fig:recursiveSchem}
\end{figure}

The exponential resource requirement can be avoided by a more sophisticated version of our algorithm that operates at a fixed resolution $\th \sim 2^{-m}$ of the oracle for arbitrary $(n,k)$.  This scalable algorithm begins by identifying candidate solutions of the number partitioning problem by searching for spin configurations in which the $m$ least significant bits of $S_z$ are zero.  To this end, we first perform Grover amplification with each coupling $J_i$ set to a value given by the $m$ least significant bits of the weights.  We thereby amplify only spin configurations for which $2^k S_z$ is a multiple of $2^m$, thus producing a superposition state with a sparser distribution of $S_z$ values than the initial state $\ket{\psi_0}$ (Fig.~\ref{fig:recursiveSchem}). We subsequently consider increasing numbers $\ell m$ of bits in successive layers $\ell = 1,2,3,\dots$ of the algorithm, setting couplings
\begin{equation}\label{eq:Jlfromw}
J_{i,\ell} = \frac{\Jmax\Mod(2^k w_i, 2^{\ell m})}{2^{\ell m}}
\end{equation}
while keeping the resolution of the oracle fixed.

This scalable algorithm retains the benefit of an efficient encoding in a central spin system, but does place additional technical demands compared with our standard algorithm.  First, the system-ancilla couplings must be changed between layers of the algorithm (App.~\ref{app:recursiveAlgorithmWeights}), a capability that is naturally present in our proposed implementation schemes.  A second new ingredient is a \textit{modular oracle} that can detect the imbalance $2^k S_z$ modulo a specified power of 2 (App.~\ref{app:recursiveAlgorithmModOracle}).  This modular oracle can readily be implemented by applying a multifrequency drive to the ancilla.  Finally, successive layers $\ell$ of the algorithm require increasingly complex operators $V_\ell$ to invert about the average amplitude of previously amplified states.  In fact, as we explain in App.~\ref{app:recursiveAlgorithmDiffusion}, the inversion step in layer $\ell$ involves repeating the entire algorithm up through layer $\ell - 1$.  For this reason, we call our scalable algorithm the \textit{recursive algorithm}.

We describe and analyze the recursive algorithm in detail in App.~\ref{app:recursiveAlgorithm},  showing that it solves the number partitioning problem in $O(2^{n/2+c n/m})$ queries, where $c = \log_2(\pi/2)$.  Thus, in the limit of a high but fixed resolution of $m\gg 1$ bits, we recover the ideal Grover speedup.  Importantly, we now attain this speedup not only in query complexity but also in the physical time to implement the algorithm in a scalable manner, in the sense that the total interaction energy required to encode the problem grows linearly with the problem size.

\begin{figure}[t]
    \centering
    \includegraphics[width=\columnwidth]{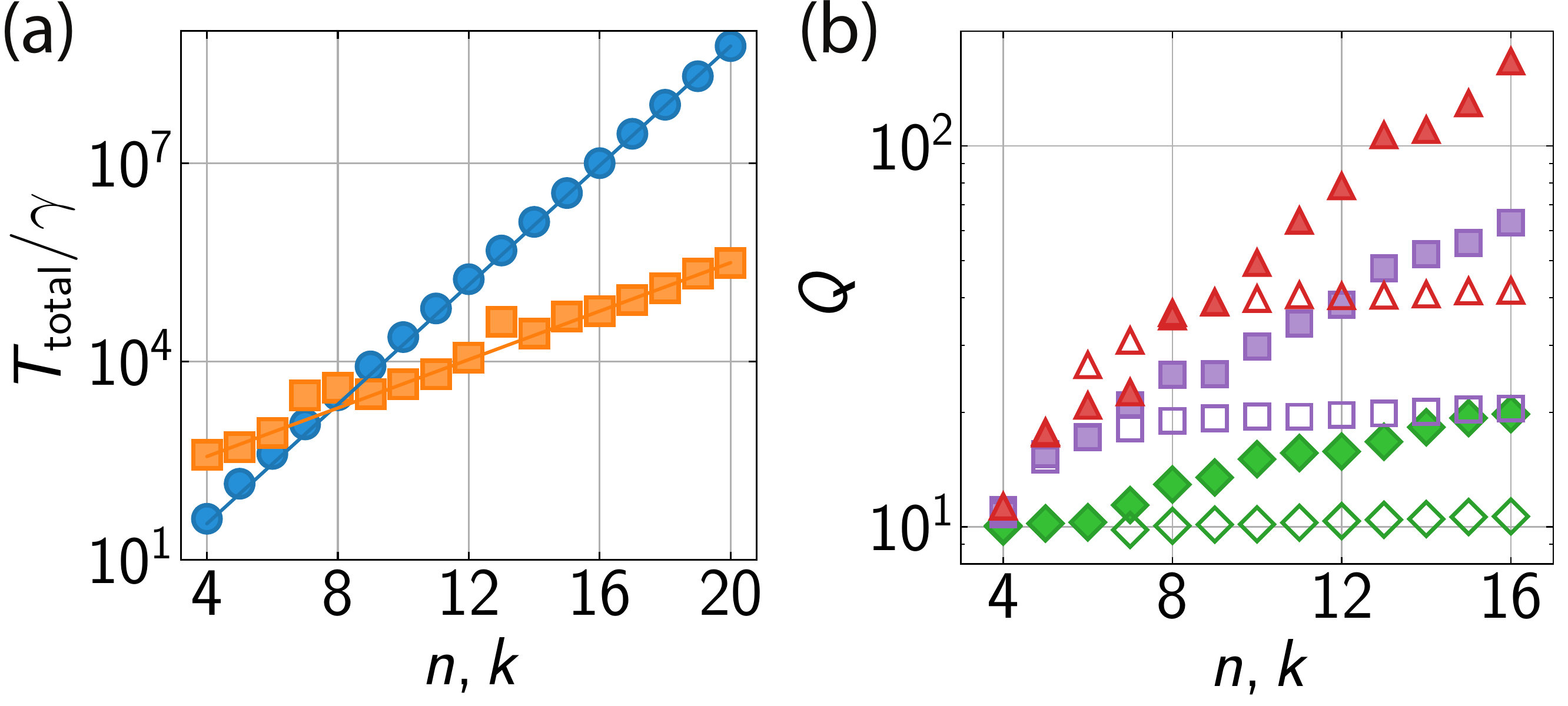}
    \caption{Comparison of the recursive and standard algorithms.
    (a) Physical time $\Ttotal/\gamma$ to implement all $\Ttotal$ queries of the oracle at fixed $\Jmax$, plotted versus $n=k$. Blue circles show the standard algorithm with $\th=\th_c$.  Orange squares show recursive algorithm with $m=6$ and $\th=2^{-m-1}$. Lines denote scalings $2^{\alpha n}$, with $\alpha = 1.5$ for the standard algorithm and $\alpha = 0.5 + 0.65/m$ for the recursive algorithm.
    (b) Speedup $Q$ versus $n=k$ for standard algorithm (open markers) and recursive algorithm (solid markers) at interaction-to-decay ratios $\rho=10^3$ (green diamonds), $\rho=10^4$ (purple squares) and $\rho=10^5$ (red triangles). 
    }
    \label{fig:recursive}
\end{figure}

A comparison of the recursive algorithm with the standard algorithm of Secs.~\ref{sec:algoimplementation}-\ref{sec:dissipation} is shown in Fig.~\ref{fig:recursive}. We simulate both algorithms with the same 1000 instances of weights to examine the scaling of their physical runtimes, which is proportional to $\Ttotal/\th$ for a fixed maximum system-ancilla coupling strength $\Jmax$ [Fig.~\ref{fig:recursive}(a)]. The physical runtime of the standard algorithm with $\th=\th_c$ scales as $\order(2^{1.5 n})$ due to the exponential narrowing of the step width $\gamma \sim 2^{-k}$ with system size $n=k$, whereas the scaling of the recursive algorithm for $m=6$ and $\th=2^{-m-1}$ is consistent with the theoretical prediction $\order(2^{0.5n + 0.65n/m})$ derived in App.~\ref{app:recursiveAlgorithm}. Thus, the recursive algorithm exhibits a scalable quantum speedup.

The performance of the recursive algorithm in realistic implementations with finite interaction-to-decay ratio shows an advantage over the standard algorithm.  In Fig.~\ref{fig:recursive}(b), we plot the speedup $\speedup$ versus $n=k$ for different interaction-to-decay ratios, with the step width $\gamma$ chosen to minimize the total number of Grover queries $\Ttotal$. In the recursive algorithm, the number of amplification cycles per layer of the algorithm (App.~\ref{app:recursiveAlgorithmDiffusion}) is additionally optimized to minimize $\Ttotal$. Whereas the speedup of the standard algorithm plateaus with increasing system size because dissipation limits the resolution of the oracle, the recursive algorithm achieves a higher speedup because it is designed to operate at fixed resolution of the oracle.

\section{Discussion and Outlook}

In this paper, we have described practical implementations of Grover's algorithm for the number partitioning problem, relying on a natural encoding in spin systems with a starlike coupling graph.  The problem offers an ideal setting for examining the physical manifestations of computational complexity, thanks to a well-understood phase diagram including easy and NP-hard regimes.  Numerical simulations of our quantum algorithm show clear signatures of the complexity phase transition, yet even in the hard phase we are able to find an advantage over an analogous classical search.

Specifically, we compared our quantum algorithm to a probabilistic classical search, with query complexity $\order(2^n)$ equivalent to that of a brute-force search (see App.~\ref{sec:classicalAlgorithms}). While there exist classical algorithms that match~\cite{horowitz1974computing, schroeppel1981algorithm} and surpass~\cite{howgrave2010new, becker2011improved} our algorithm's query complexity, they do so at the expense of exponential memory requirements~\cite{dinur2012efficient, austrin2013spacetime}. To the best of our knowledge, the leading classical algorithm of polynomial space complexity is that by Esser and May, which achieves a time complexity of $\order{\left(2^{0.645n}\right)}$~\cite{esser2019low}. Our proposed implementation achieves an improved $\order{\left(2^{0.5n}\right)}$ runtime while remaining hardware efficient, underscoring the significance of attaining a Grover speedup. Further, the possibility of using our algorithm as a subroutine in more sophisticated classical algorithms \cite{korf1998complete, schroeppel1981algorithm, becker2011improved} opens several directions for future work~\cite{bernstein2013quantum, helm2018subset, li2019improved, helm2020power, bonnetain2020improved}. 

In quantifying speedup, we have defined the runtime of the quantum algorithm in terms of the query complexity, i.e., the number of queries to the oracle.  An additional consideration is the physical time required to implement a single query.  In our standard algorithm, for a fixed maximum pairwise interaction strength $\Jmax$, the spectral resolution $\gamma\Jmax$ required of the oracle results in a query time that scales as $\gamma^{-1}\sim 2^{k_c} \approx 2^n$ along the phase boundary $k = k_c$. This exponential scaling highlights the importance of considering not only query complexity, but also the time required to implement the oracle given physical limitations of the hardware (the finite interaction energy). At finite interaction-to-decay ratio, this scaling limits the speedup of the standard algorithm in our simulations, whereas the recursive algorithm achieves higher performance limited only by the increase in optimal number of queries with system size. 



In near-term experiments, despite fragility to dissipation, even the standard algorithm could produce a speedup in few-qubit systems in the hard regime, and in scalable systems in the easy phase. In the hard regime and along the phase boundary, achieving the ideal performance at scalable system size is precluded by the exponential decrease of the energy gap with $n$. If instead we vary $n$ at fixed bit depth $k$, the time to implement each query saturates to a fixed value set by $\gamma^{-1} \sim 2^k$ as we cross the transition into the easy regime, allowing the ideal performance to be maintained at fixed interaction-to-decay ratio.  Irrespective of $k$, if we fix the duration of each query, the standard algorithm samples from a probability distribution $P(S_z)$ of fixed effective temperature set by $J_\mathrm{max}\gamma$, which may enable extensions to Boltzmann sampling~\cite{wild2020quantum}.

Our hardware-efficient approach to implementing the Grover oracle enables near-term realizations in cold-atom systems, as well as comparisons with alternative proposed methods for solving NP-hard problems in similar platforms~\cite{pichler2018quantum,zhou2020quantum}. Our approach also generalizes to other platforms, including trapped ions~\cite{figgatt2017complete} or superconducting qubits  coupled to phononic~\cite{pechal2018superconducting,hann2019hardware} or microwave~\cite{naik2017random} resonators.
The algorithms presented here might be further optimized by a variational approach that adapts the resolution of the oracle and the number of queries over multiple trials~\cite{morales2018variational}.
Grover amplification could also be applied to engineer entangled states, e.g., to produce squeezed or Dicke states by amplifying a particular $S_z$ value. For more versatile quantum control, arbitrary superpositions of Dicke states might be amplified by shaping the drive pulse~\cite{chen2015carving,keating2016arbitrary,davis2018painting}.

\begin{acknowledgments}
This work is supported by the ONR under Grant No. N00014-17-1-2279 and the AFOSR under Grant No. FA9550-20-1-0059.  O.~M. acknowledges support from the ARO under Grant No. W911NF-16-1-0490.  J.A.H., M.S.-S., and A.S.-N. acknowledge support from the DOE Office of Science, Office of Basic Energy Sciences, under Grant No. DE-SC0019174.  E.S.C. and A.P. were supported by the NSF under Grant No. PHY-1753021, the NSF GRFP (E.S.C.), and the NDSEG Fellowship (A.P.).  We thank V. Vuleti\'{c} and T.~Zhang for helpful discussions.
\end{acknowledgments}

\appendix

\section{Generalization to subset sum problem}
\label{sec:subsetsum}
The number partitioning problem is a special case of the more general class of decision problems known as subset sum problems.  These problems answer the question: given a set of $n$ objects with positive weights $w_i\in (0,1]$ of finite bit depth $k$, does there exist a subset $\mathcal{X} \subset \{w_i\}$ of total weight $\sum_\mathcal{X} w_j = \Wtarget$ for a specified value $\Wtarget$?  The entire class of problems are naturally implemented with the two experimental realizations that we present in detail in Apps.~\ref{sec:centralSpinModel}-\ref{sec:centralBosonModel}.

For general subset sum problems, implementing the oracle requires applying a $\pi$ phase shift to the system of qubits if and only if the total weight of the qubits in state $\ket{1}$ is a specified target weight $\Wtarget$, i.e., if the system is in an eigenstate of
\begin{equation}\label{eq:ss_cond}
\W \equiv \sum_i w_i \ket{1}_i\bra{1}_i
\end{equation}
with eigenvalue $\Wtarget$.  Experimentally, the target weight is set by the frequency of a field that drives the ancilla.  For the special case of the partition problem, the target weight is set to $\Wtarget = \sum_i w_i/2$, and the condition in Eq.~\ref{eq:ss_cond} then reduces to the condition $S_z\ket{x}=0$ of the main paper. More generally, the oracle phase shift in terms of $\W$ is given by
\begin{equation}\label{eq:Phi_general}
\Phi_\th (\W) = 2\arctan\left[2 (\Wtarget-\W)/\th\right] + \pi.
\end{equation}

\section{Inversion about the average}
\label{sec:inversionOperator}

The operator $V$ that performs inversion about the average, also known as the diffusion operator, requires a multiqubit controlled phase gate similar to the Grover oracle.  In particular, the operator
\begin{equation}\label{eq:diffusion}
V = 2\ketbra{\psi_0}{\psi_0} - \Id = H_n R H_n 
\end{equation}
can be decomposed into two $n$-qubit Hadamard transforms $H_n$ and a multiqubit controlled phase gate $R$~\cite{grover1997quantum}.  The operation $H_n$ is performed by applying a single-qubit Hadamard gate to each qubit. The operator
\begin{equation}\label{eq:R}
R = 2\ketbra{0}{0}-\Id
\end{equation}
is a diagonal matrix in the basis of spin configurations $\ket{x}$, with matrix elements $R_{00} = 1$ and $R_{xx} = -1$ for $x \ne 0$.  Thus, $R$ applies a phase shift of $\pi$ to all basis states except for $\ket{0}$.

The multiqubit controlled phase gate $R$ can be implemented by adapting the protocol used for the generalized oracle.  Specifically, the phase gate $R$ is equivalent, up to a global phase, to the Grover oracle for a subset sum problem (Eq.~\ref{eq:ss_cond}) with target weight zero.  A generalized version $R_\gamma$ can be implemented by setting all of the weights to the maximum value $w_i = 1$, and simultaneously choosing the detuning to set the target weight $\Wtarget = 0$.  We expect the resulting generalized diffusion operator $H_n R_\gamma H_n$ to produce the desired amplification for a relatively broad diffusion step width, requiring only $\gamma < 1$.  Notably, the step width permissible for diffusion is much broader than that required for the oracle, allowing inversion about the average to occur with negligible added dissipation even at finite interaction-to-decay ratio $\rho$. 

\begin{figure}[t]
    \centering
    \includegraphics[width=\columnwidth]{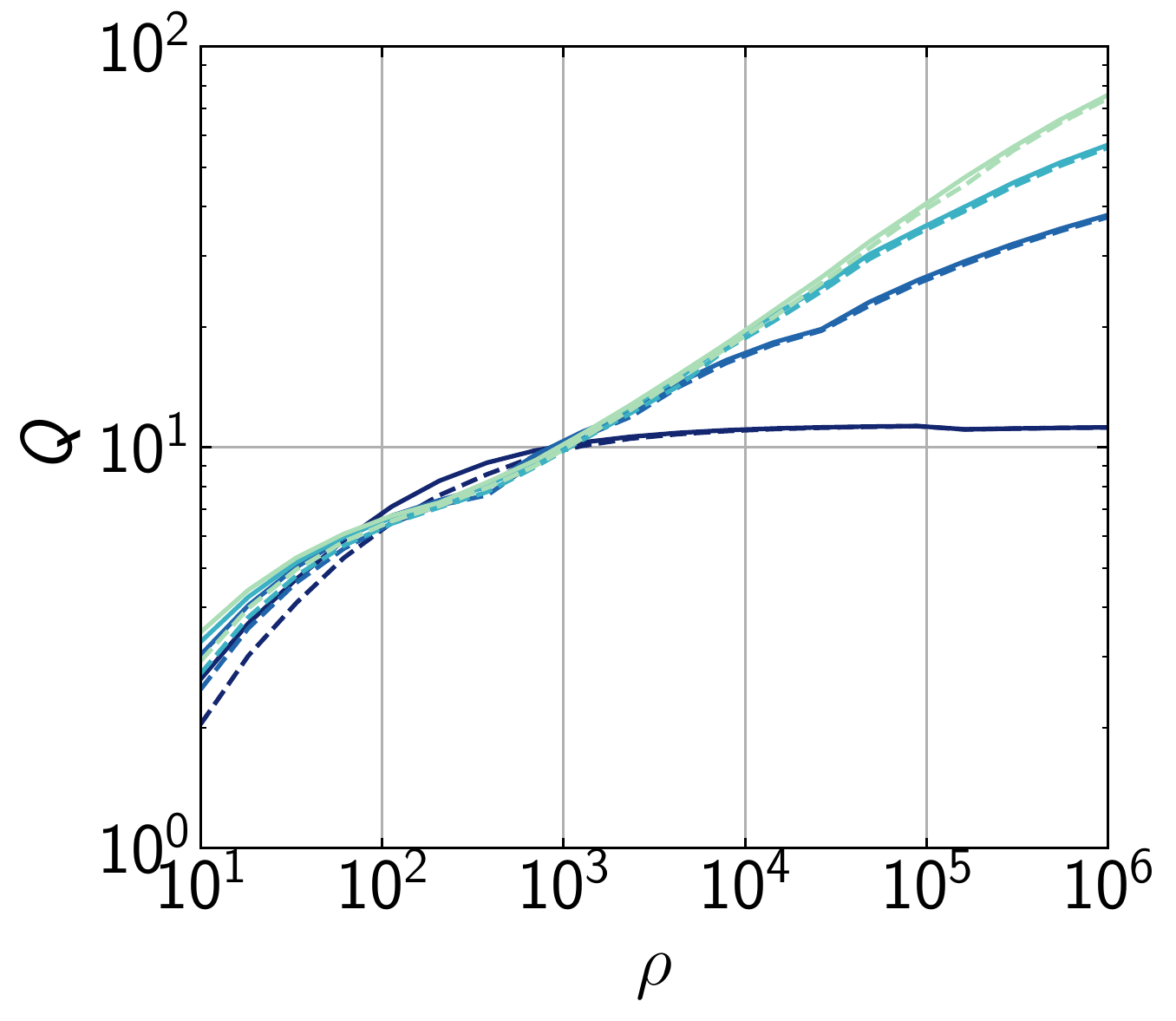}
    \caption{Comparison of speedups $Q$ between two diffusion operator methods: $Q$ versus the interaction-to-decay ratio $\rho$ for the perfect diffusion operator (full lines) and the diffusion operator using the generalized oracle (dashed lines) for $n=k=(4,6,8,10)$, shaded from darkest to lightest.
    }
    \label{fig:invOperatorSpeedupComparison}
\end{figure}

We verify that added dissipation due to the generalized diffusion operator has negligible effect by examining the quantum speedup.
In Fig.~\ref{fig:invOperatorSpeedupComparison}, we compare the achievable quantum speedup between the perfect diffusion operator and the generalized diffusion operator, in the latter case including effects of decay during diffusion as well as the nonzero step width. The speedup is reduced by at most 23\% over a wide range of $\rho$ values, thanks to the less stringent requirement on the step width during the generalized diffusion transform compared with the oracle.  Thus, for simplicity, we directly apply the ideal diffusion operator $V$ in the calculations presented in Figs.~\ref{fig:groverBlochSpheres}-\ref{fig:decay} of the main paper.

It is also possible to replace the diffusion operator with only single-qubit rotations, e.g., a global transverse field as in Ref.~\cite{jiang2017near}.  While a detailed analysis of this alternative is beyond the scope of the present work, we have simulated the application of a transverse field for a time $t=\pi/n$ in lieu of inversion about the average, finding success probabilities approximately half as large as those achieved with the multiqubit diffusion operator.  The transverse field thus enables a technically convenient scheme in which the only multiqubit gate is the oracle. 

\section{Classical search algorithms}
\label{sec:classicalAlgorithms}

In the main text, we evaluate our implementations of Grover's algorithm by comparing them to the most analogous classical algorithm, memoryless search. We begin this section by quantifying that relationship, discussing the expected and worst-case performance for each method. We then consider increasingly more complex classical number partitioning algorithms and identify their benefits and drawbacks. This allows us to consider how our algorithm compares with the \textit{best} classical algorithms, and indicates prospects for more sophisticated versions of our quantum algorithm.

Both Grover's algorithm and the classical memoryless search have a probability of success $p$ that is the same for every trial. For such search algorithms, the number of trials $M$ to obtain a solution is a random variable with expected value $E[M] = 1/p$. For Grover's algorithm the success probability is $P(\Topt) = \Popt$, as defined in the main text, so the expected number of Grover readout measurements $M_G$ is 
\begin{equation}
    E[M_G] = \frac{1}{\Popt}.
\end{equation}
For memoryless search with $N=2^n$ possible partitions and $\Nsol$ exact solutions, $p=\Nsol/N$. The expected number of memoryless trials $M_M$ is then \begin{equation}
    E[M_M] = \frac{N}{\Nsol}.
\end{equation}
Incidentally, when $T=0$, Grover's algorithm reduces to measuring an equal superposition of configuration states. The success probability is then $P_0 = \Nsol/N$, equivalent to memoryless search. 

\begin{figure}[t]
    \centering
    \includegraphics[width=\columnwidth]{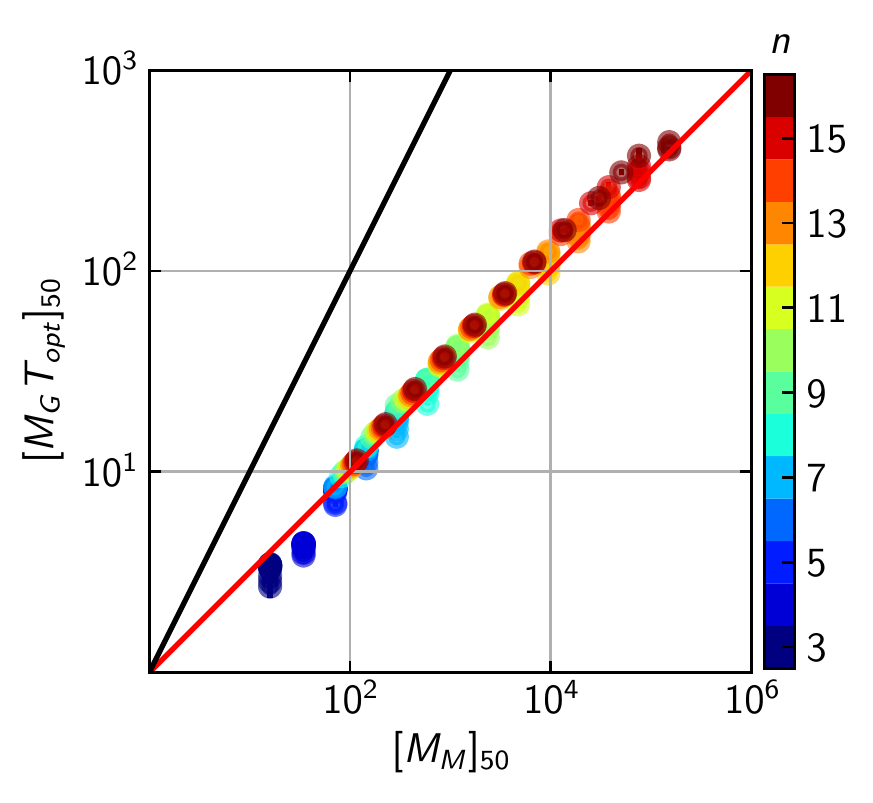}
    \caption{Speedup scaling over all $(n,k)$, 
    shown by plotting median total Grover iterations versus memoryless search trials required to reach $\Pthresh = 0.99$ probability of success. Each point represents a particular $(n,k)$, where $n$ and $k$ each take values over the range $[3, 16]$. Black and red lines denote linear and square root dependences, respectively. 
    }
    \label{fig:timeScatter}
\end{figure}

We also consider the worst-case performance of both algorithms. This is equivalent to the number of queries required to reach $\Pthresh = 1-\varepsilon$ probability of having found a solution, in the limit $\varepsilon\rightarrow 0$. For both algorithms, even after an arbitrarily large number of queries, there remains an exponentially small probability that a perfect partition exists but has not been found. We quantify this worst-case performance when $N\gg \Nsol$ by allowing $\varepsilon$ to remain finite, so that the $\Pthresh$ quantile of $M_G\Topt$ can be written as
\begin{equation}\label{eq:worst_case_grover}
    [M_G\Topt]_{\Pthresh} \in \order \left( \ln{\left(\frac{1}{\varepsilon}\right)}\times\frac{\Topt}{\Popt}\right).
\end{equation}
and the $\Pthresh$ quantile of $M_M$ is
\begin{equation}\label{eq:worst_case_memoryless}
    [M_M]_{\Pthresh} \in \order \left( \ln{\left(\frac{1}{\varepsilon}\right)}\times\frac{N}{\Nsol}\right).
\end{equation}
Figure~\ref{fig:timeScatter} shows the relative median scaling of $M_M$ and $M_G \Topt$, each calculated according to Eq.~\ref{eq:bernoulli}, with Grover's algorithm showing the expected $\sqrt{N}$ speedup.

While memoryless search follows the same probability distribution as Grover readout measurements, it is not as efficient as linear search through an unsorted list. The expected number of trials $M_L$ for linear search is
\begin{equation}
    E[M_L] = \frac{N+1}{\Nsol+1}.
\end{equation}
For $N, \Nsol \gg 1$, the expected trial scaling of both memoryless and linear search algorithms is $\order(N/\Nsol)$. The largest difference occurs with postselection in the hard regime, where $E[\Nsol]\approx2$ and memoryless search is expected to take $1.5$ times as many trials as linear search.

For the linear search, the worst-case performance is $N-\Nsol$. More generally, we can take $\varepsilon$ arbitrarily close to 0 such that $[M_L]_{\Pthresh}$ converges to $N-\Nsol$, while retaining the scaling of $[M_M]_{\Pthresh}$ in Eq.~\ref{eq:worst_case_memoryless}. Thus, while both algorithms are both worst-case linear in $N$, worst-case memoryless search requires $\order[\ln{\left(1/\varepsilon\right)}/\Nsol]$ times as many queries as worst-case linear search in the hard regime. Further, because both algorithms are unstructured, they do not need to precalculate a potentially exponential number of values before performing queries. Thus their memory scaling is $\order{\left(n\right)}$, set by the number of values to be partitioned.  

Improving upon memoryless and linear search requires us to consider a variety of structured search algorithms, which can be grouped based on the difficulty of the problem instance they aim to solve. An instance's difficulty is related to its density, defined for a set of integer weights $\mathbf{a}=\left(a_1, ..., a_n\right)$ as the ratio $d=n/\log_{2}\left(\max_{i}a_i\right)$ of the number of weights to the number of bits needed to represent the largest weight~\cite{brickel984solving, lagarias1985solving}. Thus, $d<1$ corresponds to the ``hard phase'' and $d>1$ to the ``easy phase''~\cite{mertens1998phase}. In the easy phase there are typically many perfect partitions and a problem instance can generally be solved efficiently by various classical methods, with the best based on the Karmarkar-Karp differencing algorithm~\cite{korf1998complete, Mezard2009}. In the hard phase, classical algorithms have been demonstrated to solve ``almost all'' problems of density $d<d_c<1$ in polynomial time, with subsequently published algorithms pushing $d_c$ closer to $1$~\cite{brickel984solving, lagarias1985solving, coster1992improved, schnorr1994lattice}. In such ``low-density attack'' algorithms, the number partitioning problem is reduced to the shortest vector problem, for which there exist algorithms that produce good approximations in polynomial time.

Indeed, the hardest instances of the number partitioning problem are not deep into the hard phase, but near the phase transition at a density close to 1~\cite{impagliazzo1996efficient, mertens1998phase}. For such instances, classical algorithms with the best known time complexity are subject to a space-time trade-off; improvements in runtime come at the cost of exponential memory requirements~\cite{dinur2012efficient, austrin2013spacetime}.  However, Esser and May devised a classical algorithm that achieves a time complexity of $\order{\left(2^{0.645n}\right)}$ while maintaining polynomial space complexity~\cite{esser2019low}. This algorithm offers a compelling comparison to our proposed Grover implementations, as each algorithm is hardware efficient in its use of memory or qubits. With our Grover implementation requiring $\order{\left(2^{0.5n}\right)}$ queries, a direct comparison would yield a speedup $\order{\left(2^{0.145n}\right)}$.

Finally, an interesting open question is whether one can design hardware-efficient quantum algorithms that exploit the problem structure of number partitioning. Answers to this question would build on recent work that combined quantum and classical methods to produce hybrid algorithms with exponential time, memory, and qubit trade-offs~\cite{bernstein2013quantum, helm2018subset, li2019improved, helm2020power, bonnetain2020improved}.  One avenue to explore is the use of our Grover search as a subroutine in a differencing algorithm, in which a pair of large weights $w_i, w_j$ is replaced by their difference to reduce the size of the search space.  Such differencing could be performed either classically (representing $w_i - w_j$ by a single spin) or quantumly (representing $w_i - w_j$ by an entangled state $\ket{01}+\ket{10}$ of two spins).  While classical differencing has the potential benefit of reducing the dynamic range of the weights, quantum differencing generalizes to initializing the system in a superposition state that reflects classically computed probabilities of finding certain pairs of spins on opposite sides of a perfect partition.

\section{Numerical methods}
\label{sec:numericalMethods}

The simulations of Grover's algorithm are performed numerically, by matrix multiplication according to Eq.~\ref{eq:hobbling}.  Each simulation for a specific problem size is performed on an ensemble of lists of randomly selected weights. To postselect on the existence of solutions, for each list of weights we first use the classical complete Karmarkar-Karp differencing algorithm~\cite{korf1998complete} to search for solutions, and simulate the quantum algorithm only for instances with solutions.  The number of problem instances in an ensemble, after postselection where applicable, ranges from $1000$ to $5000$ for all datasets except that used for Fig.~\ref{fig:speedup_results}(a.i), in which each probability distribution $P(S_z)$ is determined from $5\times 10^4$ instances. 

To find a sufficiently narrow step width $\th$ to reach a specified success probability $\Popt$ in Fig.~\ref{fig:speedup_results}(a.ii), we generate an ensemble of weights and numerically optimize $\th$ using the Nelder-Mead algorithm to reach the specified value $\Popt$.
To find the optimal step width $\th$ in the presence of decay [Fig.~\ref{fig:decay} and Fig.~\ref{fig:recursive}(b)], we similarly optimize $\th$ to minimize the median total number of Grover iterations using a gradient-descent algorithm.

\section{Capture range and amplification}
\label{sec:captureRange}
The interpretation of the step width $\gamma$ as a capture range for $S_z$ values is illustrated in Fig.~\ref{fig:speedup_results}(a.i) of the main text, where we plot the amplification factor after $\Topt$ Grover iterations.  Here, we additionally present an analytic derivation of the amplification factor after a single Grover iteration.  Specifically, for a given spin configuration $\ket{x}$, we show that the amplification factor after the first Grover cycle is of the Lorentzian form
\begin{equation}\label{eq:Glor}
\abs{\frac{c_{x,1}}{c_{x,0}}}^2 = \frac{A}{1+(2 S_z/\gamma)^2} + B,
\end{equation}
with width $\gamma$ set by the width of the phase step. While the amplitude $A$ and offset $B$ depend on the set of weights, we analytically derive their values averaged over instances of the weights to determine the amplification factor at $S_z=0$ as a function of step width.

We first consider the combined effect of the generalized oracle and inversion about the average on a generic state
\begin{equation}
\ket{\psi_T} = \sum_{x} c_{x,T} \ket{x}.
\end{equation}
The state $\ket{\psi_{T+1}} = VU_\gamma \ket{\psi_{T}}$ is characterized by coefficients
\begin{equation}\label{eq:iterate}
c_{x,T+1} = - e^{i\Phi_\gamma(x)}c_{x,T} + \frac{2}{N}\sum_{x'} e^{i\Phi_\gamma(x')} c_{x',T}.
\end{equation}
Equation~(\ref{eq:iterate}) simplifies for the case of $T=0$, where all coefficients $c_{x,0}=1/\sqrt{N}$ are equal.  Thus, after the first Grover iteration, we have
\begin{equation}
\frac{c_{x,1}}{c_{x,0}} = - e^{i\Phi_\gamma(x)} + \frac{2}{N}\sum_{x'} e^{i\Phi_\gamma(x')}.
\end{equation}
In terms of phasors $\z(x) = e^{i\Phi_\gamma(x)}$ and the average phasor $\zbar = \sum_x \z(x) / N$, the gain in probability of finding the system in state $\ket{x}$ is then given by
\begin{equation}\label{eq:gain}
G(x) \equiv \abs{\frac{c_{x,1}}{c_{x,0}}}^2 = 4\abs{\zbar}^2 - 4\mathrm{Re}\left[\z(x) \zbar\right] + |\z(x)|^2.
\end{equation}

We now proceed to account for the specific functional form $\Phi_\gamma(x) = 2\arctan(2S_z/\gamma)+\pi$ of the oracle's phase response.  Defining $\y(x) \equiv 2 S_z(x)/\gamma$ as the weighted spin normalized by the step width, we have
\begin{equation}
\z = \frac{\y^2-1}{\y^2+1} - i \frac{2\y}{1+\y^2}.
\end{equation}
Furthermore, since for each spin configuration $\ket{x}$ with weighted spin $S_z$ there exists a complementary spin configuration with weighted spin $-S_z$, the average phasor $\zbar$ is always real.  Equation~\ref{eq:gain} then reduces to
\begin{equation}\label{eq:Gvsmu}
G(\mu) = \left(1-2\zbar\right)^2 + \frac{8\zbar}{1 + \mu^2}.
\end{equation}
This result is of the Lorentzian form in Eq.~\ref{eq:Glor}, with amplitude $A = 8\zbar$ and offset $B = (1-2\zbar)^2$.  The gain in the first Grover cycle for a solution state ($\y = 0$) is bounded above by $\Gmax = 9$, which is achieved if $\zbar = 1$ and approached in the limit where the number of solutions is small and the step is narrow.

\begin{figure}[t]
    \centering
    \includegraphics[width=\columnwidth]{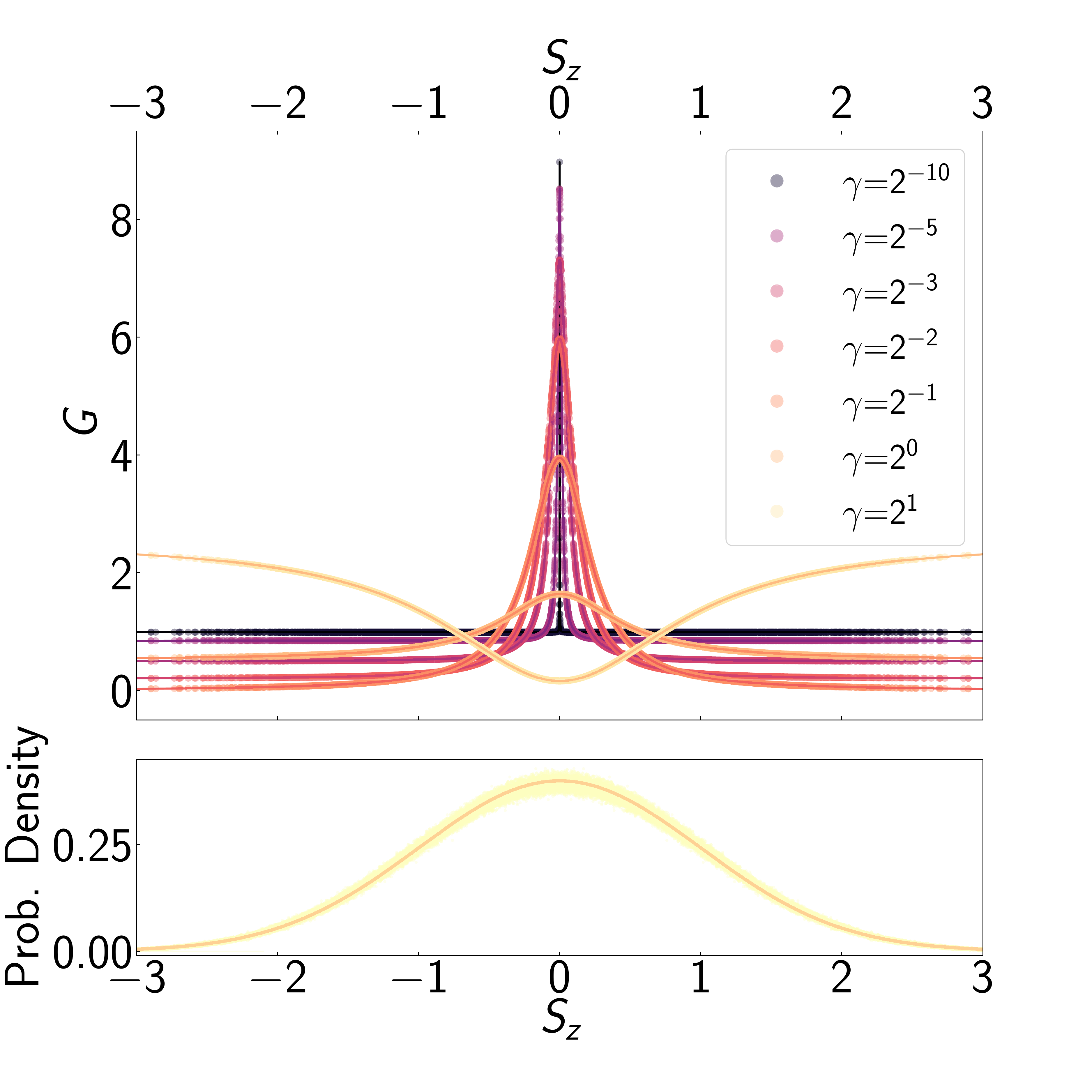}
    \caption{\textbf{Top:} Amplification factor $G$ after one Grover iteration for a single representative instance of weights at $n$=$k$=12. Fits with the Lorentzian model of Eq.~\ref{eq:Glor} are shown as solid lines. Offset $B$ is the sole free fit parameter, which is related to $A$ by Eq.~\ref{eq:Gvsmu}. \textbf{Bottom:} Initial $S_z$ probability density distribution for $n$=$k$=12, averaged over $10^4$ instances (yellow points with shading). Expected initial distribution, a Gaussian with $\sigma_{S_z} = 1$, shown as a solid orange line.
    }
    \label{fig:amplification}
\end{figure}

For illustration, we examine a single iteration of Grover's algorithm applied to number partitioning with $n=12$ random weights of bit depth $k=12$.  Figure~\ref{fig:amplification} shows the amplification factor $G$ averaged over all spin configurations $\ket{x}$ with the same value of the weighted spin $S_z$, as a function of step width $\gamma$.  Cuts at fixed $\gamma$ are well fit by the Lorentzian form in Eq.~\ref{eq:Gvsmu} with $\mu = 2S_z/\gamma$, confirming that the step width $\gamma$ sets the capture range for amplification.  The peak amplification $G_0 \equiv G(0)$ remains near its maximum possible value $\Gmax = 9$ until the width $\gamma$ grows to roughly $\sigma_{S_z}/\Gmax$, where $\sigma_{S_z}$ denotes the width of the initial $S_z$ distribution, which we plot for comparison in the bottom panel of Fig.~\ref{fig:amplification}.

The amplification $G_0$ of solution states depends to lowest order only on the ratio of $\gamma$ to the width $\sigma_{S_z}\propto \sqrt{n}$ of the $S_z$ distribution.  To calculate the dependence of $G_0$ on $\th/\sqrt{n}$ from Eq.~\ref{eq:Gvsmu}, we express $\zbar$ in terms of the number of partitions $g(\mu)$ with a given value of the imbalance $\mu$:
\begin{equation}
\zbar = \frac{1}{N}\sum_\mu g(\mu) \frac{\mu^2-1}{\mu^2+1}.
\end{equation}
Here, we have used the relation $g(\mu) = g(-\mu)$ to eliminate the term that is odd in $\mu$.  Assuming a large number $N \gg 1$ of spin configurations, we approximate the average multiplicity $\avg{g(\mu)}$ over many instances of the weights using a normal distribution
\begin{equation}
p(\mu) = \frac{1}{\sigma\sqrt{2\pi}} e^{-\mu^2/(2\sigma^2)}
\end{equation}
of standard deviation $\sigma = w_\mathrm{rms}\sqrt{n}/\gamma$, where $\wrms \equiv \sqrt{\avg{w_i^2}}=1/\sqrt{3}$ for weights chosen from a uniform distribution on $(0,1]$. In terms of $p(\mu)d\mu \approx \avg{g(\mu)}/N$, we have
\begin{align}\label{eq:gaussian_integral}
\avg{\zbar} &= \int_{-\infty}^\infty p(\mu) \frac{\mu^2-1}{\mu^2+1} \, d\mu\nonumber\\
&= 1- \frac{\sqrt{2\pi}e^{1/(2\sigma^2)}}{\sigma}\mathrm{erfc}\left(\frac{1}{\sqrt{2}\sigma}\right),
\end{align}
where $\mathrm{erfc}$ is the complementary error function.

The average amplification over many instances of the weights is given in terms of $\zbar$ by
\begin{equation}\label{eq:G0avg}
\avg{G_0} = 1 + 4\avg{\zbar} + 4\avg{\zbar^2}\geq 1 + 4\avg{\zbar} + 4\avg{\zbar}^2.
\end{equation}
This bound is tight in the large-$N$ limit, where the variance in $\zbar$ over different instances of the weights is small. We plot the lower bound in Eq.~\ref{eq:G0avg} as the dashed red curve in Fig.~\ref{fig:decay}(c).  There, we denote the amplification as $Q_1 \equiv G_0$ to emphasize its equivalence to the quantum speedup for a single Grover cycle.  We compare our model with the amplification calculated at $n=k$ for $n=12$, in each case averaging over $10^3$ instances of the weights with postselection.  We observe excellent agreement between the model and the simulation.

\section{Scalable algorithm}
\label{app:recursiveAlgorithm}

In Sec.~\ref{sec:recursive}, we outline a recursive version of our algorithm that allows for operating at a fixed resolution $\gamma \sim 2^{-m}$ of the oracle for arbitrary problem size.  The essence of our approach is to consider only the $\ell m$ least significant bits of the weights for successive values $\ell = 1, 2, 3, \dots$.  These truncated weights suffice to amplify, in each layer $\ell$ of the algorithm, candidate solutions satisfying the condition $\tmod(2^k S_z, 2^{\ell m}) = 0$.  In the final layer of the algorithm, the fixed $m$-bit resolution of the oracle suffices to identify only true solutions satisfying $S_z = 0$, thanks to the preamplification of a sparse distribution of $S_z$ values in prior layers.

In this appendix, we elaborate on the details of the scalable algorithm, including the encoding of the weights, the modular oracle, and the recursive implementation of inversion about the average.  Finally, we derive the asymptotic scaling of the query complexity and present numerical simulations corroborating our analysis.

\subsection{Encoding the weights}\label{app:recursiveAlgorithmWeights}
Key to our approach is the ability to dynamically change the mapping from weights $w_i$ to system-ancilla couplings $J_i$ between successive queries of the oracle. We define the following set of mappings from weights to system-ancilla couplings:
\begin{equation}\label{eq:Jfromw}
J_{i,\ell} = \frac{\Jmax\Mod(2^k w_i, 2^{\ell m})}{2^{\ell m}},
\end{equation}
with $\ell = 1,2,\dots k/m$.  Here, we scale the couplings to a fixed maximum value $\Jmax$ as usual, but for a given value $\ell$ we use only the $\ell m$ least significant bits of the weights.  (For $\ell > 1$, we are keeping more bits than the oracle can actually resolve, to avoid subtleties of accounting for carry bits that can add up.  It should never be necessary to program the weights with a resolution of more than $m + \log_2(n)$ bits, but the higher precision assumed in Eq.~\ref{eq:Jfromw} also does no harm.)

\subsection{Grover amplification with modular oracle}
\label{app:recursiveAlgorithmModOracle}

In each layer $\ell$ of our algorithm, our objective is to amplify spin configurations satisfying the condition $\tmod(2^k S_z, 2^{\ell m}) = 0$, i.e., spin configurations for which the $\ell m$ least significant bits of the imbalance $S_z$ are zero.  To check this condition for a given value of $\ell$, we need only to know the $\ell m$ least significant bits of each weight, so we use the couplings $J_{i,\ell}$ defined in Eq.~\ref{eq:Jfromw} for each layer $\ell$.  We further define

\begin{align}
S_{z,\ell} &= 2^{-\ell m} \sum_{i=1}^n \Mod(2^k w_i, 2^{\ell m})\sigma^z_i/2 \nonumber \\
&= \frac{1}{\Jmax}\sum_{i=1}^n J_{i,\ell}\sigma^z_i/2,
\end{align}
representing the imbalance at layer $\ell$ using only the $\ell m$ least significant bits of the weights.  We wish to design the oracle to produce a $\pi$ phase shift if $\tmod(2^{\ell m} S_{z,\ell},2^{\ell m}) = 0$, which is equivalent to the ancilla resonance energy shift being an integer multiple of $\Jmax$.

This modular oracle can be implemented in the central spin or central boson model by subjecting the ancilla to a multifrequency drive field, consisting of a comb with spacing $\Jmax$.  Since $S_{z,\ell} \leq n$, there are $\sim n$ different possible values $2^{\ell m} S_{z,\ell}$ that are equivalent to zero modulo $2^{\ell m}$, and correspondingly only approximately $n$ drive frequencies are needed.  Each tooth of the comb of drive fields has a spectral width $\th\Jmax$, which we will choose to be independent of $\ell$, with a value $\th \sim 2^{-m} \ll 1$.  The separation of scales between the width and the spacing of the teeth ensures that the phase response of the oracle is well approximated as
\begin{equation}\label{eq:Phil}
\Phi_\ell \approx 2\arctan\left[\frac{2\Mod(2^k S_z,2^{\ell m},-2^{\ell m -1})}{2^{\ell m}\th} \right] + \pi,
\end{equation}
where $\tmod(\cdot,d,b)$ denotes the modulo with divisor $d$ and offset $b$.  In terms of the phase shift $\Phi_\ell$ at layer $\ell$, we define the modular oracle $U_\ell = \exp(i \Phi_\ell)$. In the final layer of the algorithm, where $\ell m=k$, we want to amplify only the partitions with $S_z=0$ without taking the modulus, so we apply our usual oracle $U_{k/m} = \exp(i \Phi)$ with resolution $\th$.

\subsection{Recursive algorithm}
\label{app:recursiveAlgorithmDiffusion}

The first layer of our algorithm consists simply of applying the modular oracle in alternation with the diffusion operator $V = H_n R H_n$, where $H_n$ is the $n$-qubit Hadamard and $R$ is a multiqubit controlled phase gate  (App.~\ref{sec:inversionOperator}).  For an imperfect oracle, we apply the usual spin-echo sequence to produce a state
\begin{equation}\label{eq:psi1}
\ket{\psi_1}  = (V U_1^\dagger V U_1)^{T_1/2} \ket{\psi_0}.
\end{equation}
where $\ket{\psi_0} = H_n\ket{0}^{\otimes n}$ is the equal superposition of all spin configurations.  We assume the number of amplification cycles $T_1$ to be even for notational simplicity, but Eq.~\ref{eq:psi1} can also be generalized to allow an odd number of cycles.  Since only a fraction $2^{-m}$ of the spin configurations satisfy the condition $\tmod(2^m S_{z,1},2^m) = 0$, we expect to require approximately $T_\ell \approx (\pi/4)2^{m/2}$ amplification cycles in the first layer $\ell = 1$.  Upon completion of this layer of the algorithm, the state $\ket{\psi_1} \equiv \mathcal{G}_1 \ket{\psi_0}$ is approximately an equal superposition of all spin configurations that are candidate solutions to the number partitioning problem based on the $m$ least significant bits of $S_z$.

Naively one might expect subsequent layers of our algorithm to be analogous to Eq.~\ref{eq:psi1} with the replacement $U_1 \rightarrow U_\ell$.  However, an important subtlety is that the diffusion operator $V$ must be modified so that at layer $\ell$ it rotates by $\pi$ about the state $\ket{\psi_{\ell-1}}$, i.e., 
\begin{equation}
    V_\ell=2\ketbra{\psi_{\ell-1}}{\psi_{\ell-1}}-\Id. 
\end{equation}
In particular, it is important to rotate about $\ket{\psi_{\ell-1}}$ --- as opposed to $\ket{\psi_0}$ --- so that the amplitude of the solution states is inverted about the average amplitude in the sparse superposition of $S_z$ values produced in the preceding layer, while ignoring the near-zero amplitudes of the spin configurations that have already been suppressed.

To understand how to implement the generalized diffusion operator $V_\ell$, we first recall how our usual diffusion operator is constructed (Eqs.~\ref{eq:diffusion}-\ref{eq:R}).  We can perform a $\pi$ rotation about any state $\ket{\psi}$ by a combination of (1) the operator $R$ that rotates about the state $\ket{0}\equiv \ket{0}^{\otimes n}$ and (2) a unitary operator $\mathcal{O}$ that transforms $\ket{\psi}$ to $\ket{0}$.  In terms of these ingredients, the rotation about $\ket{\psi}$ is implemented by applying the compound operator $\mathcal{O}^\dagger R \mathcal{O}$.  For the usual Grover’s algorithm, the $n$-qubit Hadamard $\mathcal{O} = H_n = \mathcal{O}^\dagger$ is the operator that transforms $\ket{\psi}$ to $\ket{0}$ and back.

To construct the diffusion operator $V_\ell$ for any layer of our algorithm, we thus require an operator $\mathcal{O}$ that transforms the state $\ket{\psi_{\ell-1}}$ to state $\ket{0}$.  Conveniently, we know exactly how to perform this transformation for arbitrary $\ell$ --- by applying Grover’s algorithm all the way up to layer $\ell - 1$.  If we define the Grover operator at level $\ell$ as
\begin{equation}
\mathcal{G}_\ell \equiv \left(V_{\ell} U_\ell^\dagger V_\ell U_\ell\right)^{T_\ell/2},
\end{equation}
such that $\ket{\psi_\ell} = \mathcal{G}_\ell \ket{\psi_{\ell-1}}$, then the operator $\mathcal{O}^\dagger=\left(\prod_{\ell'=1}^{\ell - 1}\mathcal{G}_{\ell'}\right) H_n$ transforms $\ket{0}^{\otimes n}$  to $\ket{\psi}_{\ell-1}$.
Thus, the diffusion operator needed in layer $\ell$ of the algorithm is
\begin{equation}\label{eq:Vl}
V_\ell =  \left(\prod_{\ell'=1}^{\ell - 1}\mathcal{G}_{\ell'}\right) H_n R H_n \left(\prod_{\ell'=1}^{\ell - 1}\mathcal{G}_{\ell'}\right)^\dagger .
\end{equation}
Note that Eq.~\ref{eq:Vl} correctly reduces to $V_1 = V$ for the first layer of our algorithm.

\subsection{Query complexity}\label{app:query_complexity}
Due to the recursive nature of the algorithm, the query complexity grows exponentially with $k$ and hence with $n$ in the hard regime.  This should not surprise us, since Grover's algorithm cannot produce an exponential speedup.  The key performance metric, then, is the coefficient $\alpha$ in the exponent of the $\order(2^{\alpha n})$ query complexity.

The query complexity is given by
\begin{equation}\label{eq:TtotSeries}
T_\mathrm{tot} = \sum_{\ell =1}^{k/m} T_\ell \tau_\ell
\end{equation}
where $T_\ell$ is the number of amplification cycles at layer $\ell$ and $\tau_\ell$ is the number of calls to the oracle required in each amplification cycle, including the queries involved in implementing the diffusion operator $V_\ell$ for $\ell > 1$.  We expect to need $T_\ell \approx (\pi/4)2^{m/2}$ amplification cycles at each layer except the final one, by the same argument given above for $\ell = 1$. The final layer takes a factor of $\sqrt{n}$ more steps, but this factor will only introduce a subexponential correction to the query complexity so we can ignore it in the following analysis.  The number of calls to the oracle in each amplification cycle of the $\ell^\mathrm{th}$ layer is
\begin{equation}
\tau_\ell = 1+\sum_{\ell'=1}^{\ell-1}2T_{\ell'}\tau_{\ell'},
\end{equation}
based on Eq.~\ref{eq:Vl}.  Put another way, we have
\begin{align}
\tau_\ell &= \tau_{\ell-1}\left(1 + 2T_{\ell-1}\right) \nonumber \\
&\approx \tau_{\ell-1}\left[1 + 2^{m/2} \left(\frac{\pi}{2}\right)\right].
\end{align}
Since the first layer requires only $\tau_1 = 1$ call to the oracle per amplification cycle, for general $\ell$ we have
\begin{equation}
\tau_\ell = \left[1 + 2^{m/2} \left(\frac{\pi}{2}\right)\right]^{\ell -1},
\end{equation}
as can readily be verified by induction.

The total number of calls to the oracle given by Eq.~\ref{eq:TtotSeries} thus takes the form of a finite geometric series.  Evaluating the geometric series yields
\begin{align}\label{eq:Ttot}
T_\mathrm{tot} &= \left(\frac{\pi}{4}\right)2^{m/2}\left(\frac{\left[1 + 2^{m/2} \left(\frac{\pi}{2}\right)\right]^{k/m} - 1}{2^{m/2} \left(\frac{\pi}{2}\right)}\right) \nonumber \\ 
&\approx 2^{k/2-1}\left(\pi/2 \right)^{k/m} \nonumber \\
&= 2^{k(1/2 + c/m)-1},
\end{align}
where $c = \log_2(\pi/2) \approx 0.65$.  For $n\approx k$, we obtain a query complexity $\order(2^{\alpha n})$ with $\alpha = 0.5 + 0.65/m$.  We thus need $m\approx 5$ bits of resolution to outperform the best scalable classical algorithm \cite{esser2019low}.

\begin{figure*}[t] 
    \centering
    \includegraphics[width=\textwidth]{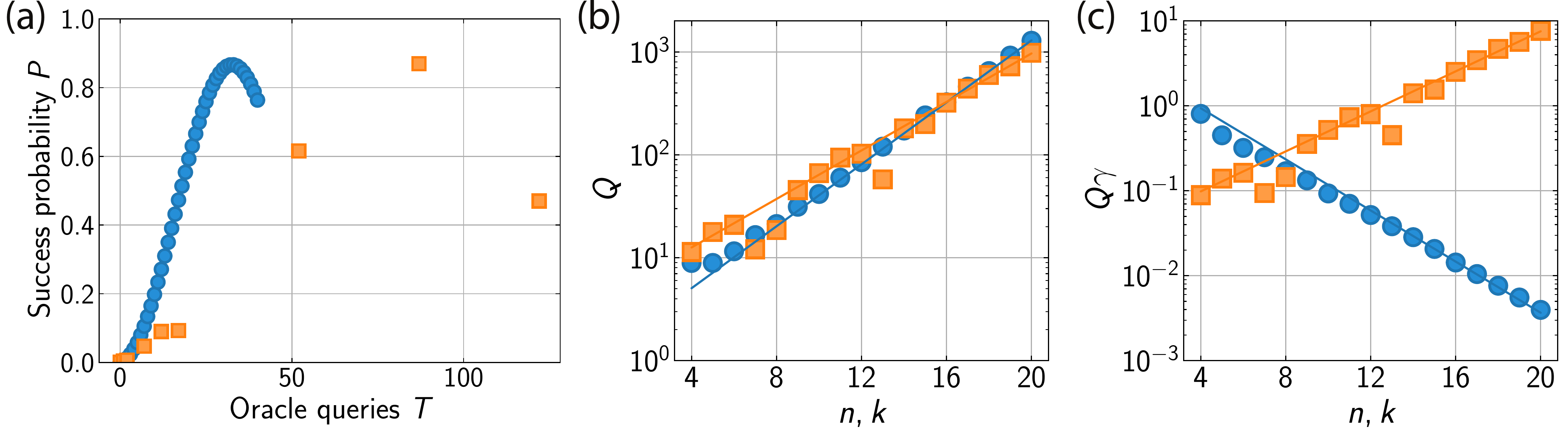} 
    \caption{Comparison of the standard (blue circles) and recursive (orange squares) algorithms. (a) Success probability $P$ versus number of oracle queries $T$ for a single instance of number partitioning at $n=k=12$.  Recursive algorithm is performed with $m=4$ and $\th=2^{-m-1}$.  The physical time per query (at fixed $\Jmax$) is longer by a factor of $2^{k-m-1}=2^7$ for the standard algorithm than for the recursive algorithm. (b) Median speedup $\quantile{Q}{0.5}$ in query complexity, plotted versus $n=k$ for the standard algorithm with $\th=\th_c$ and the recursive algorithm with $m=6$ and $\th=2^{-m-1}$. Lines denote the $2^{0.5n}$ scaling for the standard algorithm and $2^{(0.5-0.65/m)n}$ scaling for the recursive algorithm.
    (c) Speedup $\quantile{Q}{0.5} \gamma$ in physical runtime at fixed $\Jmax$, plotted versus $n=k$ for the standard algorithm with $\th=\th_c$ and the recursive algorithm with $m=6$ and $\th=2^{-m-1}$. Lines denote the $2^{-0.5n}$ scaling for the standard algorithm and $2^{(0.5-0.65/m)n}$ scaling for the recursive algorithm.
    }
    \label{fig:recursiveAlgorithmComparison}
\end{figure*}

While the query complexity of the recursive algorithm is at best (i.e., for large $m$) the same as that of the standard algorithm, the recursive algorithm offers the benefit that the actual runtime in a scalable implementation with fixed $\Jmax$ is directly proportional to the query complexity, and thus exhibits a Grover speedup.  We thus eliminate the exponential overhead that is present in the simplest algorithm, and we have do so without compromising on hardware efficiency.

\subsection{Simulation}

A representative comparison of the standard algorithm and the recursive algorithm is shown in Fig.~\ref{fig:recursiveAlgorithmComparison}(a), where we simulate a single instance of number partitioning with $n=k=12$.  The instance is selected to have exactly one pair of solutions.  The recursive algorithm was performed with $m=4$ bits of resolution, and the amplification steps per layer $(T_1, T_2, T_3) = (2,3,2)$ are chosen to maximize the probability at each layer.  The resolution of the oracle is set to $\gamma = 2^{-m-1}$ for the recursive algorithm (orange squares), compared with $\gamma = 2^{-k}$ for the standard algorithm (blue circles).  While the recursive algorithm required approximately 3 times as many queries as the standard algorithm, the physical time per query at fixed $\Jmax$ is a factor of $2^{k-m-1}=2^7$ times longer for the standard algorithm than for the recursive algorithm.  Thus, in this example the recursive algorithm produces a significant reduction in runtime for a fixed maximum system-ancilla coupling.

To compare the time complexity of the algorithms, we simulate both algorithms for a range of problem sizes $(n,k)$ with 1000 instances of weights [Fig.~\ref{fig:recursiveAlgorithmComparison}(b,c)]. For each layer $\ell$ of the recursive algorithm with $m=6$ and $\th = 2^{-m-1}$, the number of amplification cycles $T_\ell$ is optimized to minimize the median total number of Grover oracle queries $\Ttotal$. We expect the total number of Grover queries $\Ttotal$ in the recursive algorithm to follow the query complexity derived in the preceding section (App.~\ref{app:query_complexity}). For $m=6$, the expected scaling is $\Ttotal \in \order(2^{0.61 n})$, leading to a $\speedup \in \order(2^{0.39 n})$ scaling of the speedup with system size, which is confirmed by the simulation in Fig.~\ref{fig:recursiveAlgorithmComparison}(b). While the speedup $Q$ in query complexity for the recursive algorithm at finite bit depth $m$ is slightly lower than that of the standard algorithm, the benefit of the recursive algorithm becomes apparent when we plot the speedup $Q\gamma$ in physical runtime at fixed $\Jmax$ [Fig.~\ref{fig:recursiveAlgorithmComparison}(c)].  The growth in $Q\gamma$ with system size in the recursive algorithm confirms its scalability.

\section{Effects of decoherence}
\label{app:dissipation}

Two forms of decoherence that can limit the performance of our algorithm in realistic implementations are decay of the ancilla and decay of the system spins.  In this section, we first provide an analytic estimate of the scaling of the quantum speedup with a generic interaction-to-decay ratio in the standard algorithm (App.~\ref{sec:speedupAndDecay}).  We then describe how we calculate the speedup in the numerical simulations of Fig.~\ref{fig:decay}, focusing on decay of the ancilla, which is the dominant decay channel in the near-term experimental implementations proposed and analyzed in App.~\ref{app:implementations}.

\subsection{Quantum speedup in presence of decay}
\label{sec:speedupAndDecay}

\begin{figure}[t]
    \centering
    \includegraphics[width=\columnwidth]{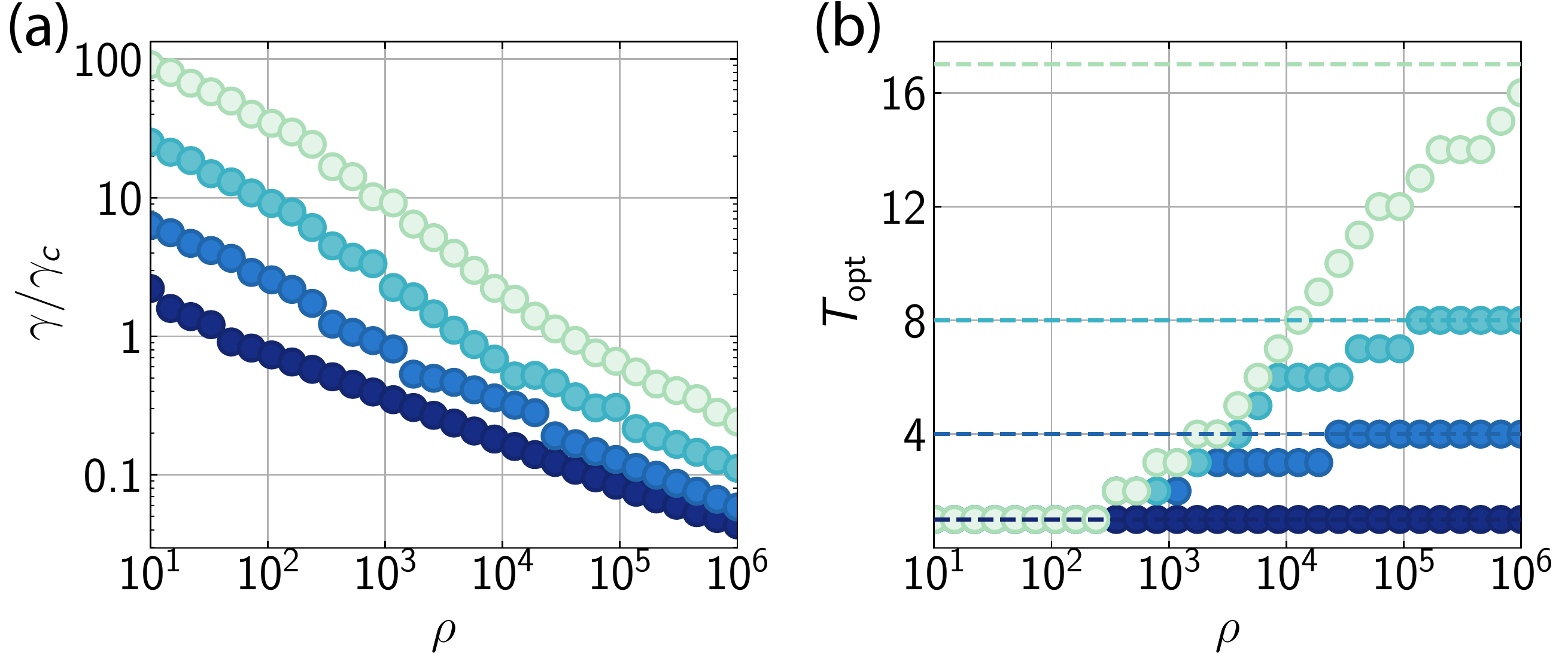}
    \caption{
    (a) Ratio of optimal step width $\th$ to critical step width $\th_c$ versus the interaction-to-decay ratio $\rho$ for $n=k=(4,6,8,10)$ denoted by markers shaded from darkest to lightest.
    (b) Optimal number of Grover iterations $\Topt$ versus the interaction-to-decay ratio $\rho$ for $n=k=(4,6,8,10)$ denoted by markers shaded from darkest to lightest. Dashed lines represent the number of iterations $\Topt$ at which the speedup is maximized in the ideal Grover's algorithm.
    }
    \label{fig:stepWidthToptVsRho}
\end{figure}

Decay during the generalized Grover's oracle limits the maximum achievable quantum speedup. Here, we analytically derive the scaling of optimal quantum speedup with the interaction-to-decay ratio for the standard algorithm presented in Secs.~\ref{sec:algoimplementation}-\ref{sec:speedup}. The speedup is maximized at a step width $\thopt$ set by a competition between the reduction in capture range at narrower step widths, which ideally increases the success probability, and the accompanying increase in decay. Figure~\ref{fig:stepWidthToptVsRho} shows the optimal step width and the optimal number of Grover iterations $\Topt$ that produce the speedup shown in Fig.~\ref{fig:decay}(b) of the main text. At small interaction-to-decay ratios, it is optimal to use a single amplification cycle with a wide phase step, while at larger interaction-to-decay ratios, the optimal step width is narrower, allowing for a performance closer to that of the ideal Grover's algorithm.

To estimate the optimal step width, we observe that the number of partitions $\Neff(\th)$ within the capture range $\abs{S_z}\lesssim \gamma$ sets the behavior of the generalized Grover's algorithm in roughly the same way as the number of perfect partitions $\Nsol$ sets the behavior of the ideal Grover's algorithm.  With increasing step width, in the absence of dissipation, the number of iterations required to maximize the success probability decreases as
\begin{equation}
    \Topt^* \approx \frac{\pi}{4} \sqrt{\frac{N}{\Neff}},
\end{equation}
in analogy to Eq.~\ref{eq:Tgrover}.  (In defining $\Topt^*$ to maximize the success probability, we are choosing a slightly different definition from that of $\Topt$ in the main text.)  For large step widths, where we capture a larger number $\Neff$ of spin configurations than the actual number of solutions $\Nsol$, we can approximate $\Neff \approx \th N \sqrt{\frac{6}{\pi n}}$ from the theoretical distribution of total weights in the partition problem~\cite{mertens2000random, mertens2006easiest}.  Thus, in tems of the step width $\gamma$, we have
\begin{equation}\label{eq:Toptgamma}
\Topt^* = \frac{\pi}{4\gamma^{1/2}}\left(\frac{\pi n}{6}\right)^{1/4}.
\end{equation}

The decrease in the optimal number of iterations $\Topt^*$ with increasing step width comes at the cost of a reduced success probability $\Popt \approx \Nsol/\Neff$, even before accounting for dissipation.  Thus, employing a narrower step for a larger number of iterations $T$ is preferable unless decay results in an appreciable reduction in $\Popt$.  To estimate the optimal number of Grover iterations at finite interaction-to-decay ratio $\rho$, we first determine the maximum number $T_C$ of iterations that can be performed with a given probability $e^{-C}$ of incurring no error. Here, $C$ is a constant that we choose to optimize the speedup. The error rate per iteration is $D/(\rho\gamma)$, where $D$ is an order-unity factor that is derived in App.~\ref{sec:decoherence} for the case of the first amplification step and, more generally, can be obtained from a fit to numerical data.  We thus estimate the maximum number of iterations as $T_C \approx C\rho\th/D$.

We expect the optimum number of iterations in the presence of decay to be given by $\Topt^* = T_C$ for some order-unity value $C$.  Combining the expression for $T_C$ and the relationship between $\Topt^*$ and $\thopt$ (Eq.~\ref{eq:Toptgamma}), the optimal step width is then
\begin{equation}
    \thopt = \left ( \frac{\pi D}{4 C \rho}\right )^{2/3} \left ( \frac{\pi n}{6}\right )^{1/6}.
\end{equation}

To estimate the speedup $\speedupopt$, we approximate $\Popt$ in the presence of dissipation as $\Popt \approx e^{-C} \Nsol/\Neff $. The speedup $\speedupopt$ is then given by
\begin{equation}
    \speedupopt = \frac{\log(1-\Popt)}{\Topt^* \log(1-P_0)} \approx \frac{\Popt}{\Topt^* P_0},
\end{equation}
where $P_0=\Nsol/N$ and we assume $\Popt \ll 1$ and $P_0 \ll 1$.
Finally, collecting the expressions, we find 

\begin{align}
 \speedupopt &= \frac{e^{-C}}{\Topt^* \thopt}  \sqrt{\frac{\pi n}{6}} \nonumber \\
        &=  \left(\frac{4}{\pi} \right)^{4/3}\left (\frac{\pi n}{6} \right)^{1/6}e^{-C} \left(\frac{C \rho}{D}\right)^{1/3}.
\label{eq:speedupDecayScaling}
\end{align}

The scaling of the optimal speedup as a function of interaction-to-decay ratio is given by $\speedupopt \sim \rho^{1/3}$. For high values of $\rho$, the optimal speedup will start to saturate to the quantum speedup of the ideal Grover's algorithm. This saturation occurs when the optimal step width becomes smaller than the smallest nonzero $\abs{S_z}$ values, which for $n=k$ is at $\thopt \approx \sqrt{n}/N$, with $N=2^n$. Thus, the interaction-to-decay ratio where the speedup starts to saturate scales as $\rho \sim N^{3/2}/\sqrt{n}$. This scaling exemplifies the fact that reaching the ultimate quantum speedup allowed by Grover's algorithm requires exponentially increasing the interaction-to-decay ratio with problem size.

The numerical results of the generalized Grover's algorithm with ancilla decay in Fig.~\ref{fig:decay}(b) are well described by the model of Eq.~\ref{eq:speedupDecayScaling} with constants $C=1/3$ and $D=1.2$. This equation is applicable in a region between $100 \lesssim \rho \lesssim N^{3/2}/\sqrt{n}$. The upper limit of this regime of validity comes from the saturation of the speedup to the ideal Grover's algorithm limit, while the lower limit is reached when $\Topt^* = 1$. 

\subsection{Generalized oracle with ancilla decay}
\label{sec:decoherence}

The effect of ancilla decoherence during the generalized Grover's oracle can be modeled as an imaginary term in the oracle phase shift (Eq.~\ref{eq:Phi}). A particular system spin configuration $\ket{x}$ will shift the ancilla excited state from resonance by $\Delta_x = (\Wtarget - \W)\Jmax$, where $\W$ and $\Wtarget$ are the actual and target weights in the subset sum problem as defined in Appendix~\ref{sec:subsetsum}. To include the effect of ancilla decoherence, we make a substitution $\Delta_x \xrightarrow{} \Delta_x + i \Gamma_a/2$, where $\Gamma_a$ is the linewidth of the ancilla excited state~\cite{cohen1992atom}. Thus, the oracle phase shift applied to the spin configuration $\ket{x}$ is given by
\begin{align}
    \Phi_{\gamma} (\W) &= 2\arctan\left[2(\Wtarget-\W)/\th + i \Gamma_a/ ( \Jmax \th) \right] + \pi \nonumber \\
    &= 2\arctan\left(\mu + i \rr \right) + \pi.
\label{eq:PhiDecay}
\end{align}
Here, $\mu=2(\Wtarget-\W)/\th$ in an analogy to the definition in App.~\ref{sec:captureRange} and
\begin{equation}
\rr\equiv\frac{\Gamma_a}{\Jmax\gamma} = \frac{1}{\rho\gamma}
\end{equation}
parameterizes the decay rate per query of the oracle, assuming the decay is dominated by the ancilla decay.

The effect of the oracle on the amplitudes of the spin states is given by $\z(\W)=\exp[i\Phi_{\gamma} (\W)]$. Using Eq.~\ref{eq:PhiDecay} we derive
\begin{equation}
    \z(\W) = -\frac{1+i \mu - \rr}{1-i \mu + \rr}.
\label{eq:zDecay}
\end{equation}
This full form of the oracle including dissipation modifies the single-cycle amplification formula given in App.~\ref{sec:captureRange}. To see how, we rewrite $\z$ in terms of its real and imaginary components,
\begin{equation}
\z = \frac{\y^2+\rr^2-1}{\y^2+(\rr+1)^2} - i \frac{2\y}{(1+\rr)^2+\y^2},
\end{equation}
where we use the fact that both $\rr$ and $\y$ are real.

The expression for the amplification in Eq.~\ref{eq:gain} now reduces to
\begin{equation}\label{eq:Gvsmu_dis}
G(\mu) = 4\zbar\left(\zbar-1\right) + \frac{(1-\rr)^2}{(1+\rr)^2} +\frac{8\zbar(1+\rr)}{(1 + \rr)^2 + \mu^2}.
\end{equation}
As before, $\zbar$ is real and thus depends only the real components of $\z$, weighted by the density of states $g(\mu)$:
\begin{equation}
\zbar = \frac{1}{N}\sum_\mu g(\mu) \frac{\mu^2+\rr^2-1}{\mu^2+(\rr+1)^2}.
\end{equation}
Taking the continuum limit and using the probability distribution $p(\mu)$ derived in App.~\ref{sec:captureRange} yields the updated expectation value,
\begin{align}\label{eq:chibar_diss}
\avg{\zbar} &= \int_{-\infty}^\infty p(\mu) \frac{\mu^2+\rr^2-1}{\mu^2+(\rr+1)^2} \, d\mu\nonumber\\
&= 1- \frac{\sqrt{2\pi}e^{(1+\rr)^2/(2\sigma^2)}}{\sigma}\mathrm{erfc}\left(\frac{1+\rr}{\sqrt{2}\sigma}\right).
\end{align}
From Eqs.~(\ref{eq:Gvsmu_dis}) and~(\ref{eq:chibar_diss}) we compute the average amplification over many instances:

\begin{align}\label{eq:G0avg_dis}
\avg{G_0} \geq \frac{(1-\rr)^2}{(1+\rr)^2} + \left( \frac{8}{1+\rr}-4\right)\avg{\zbar} + 4\avg{\zbar}^2.
\end{align}

The amplification in Eq.~\ref{eq:G0avg_dis} is a lower bound both due to the substitution of $\avg{\zbar}^2$ for $\avg{\zbar^2}$ and due to the small additional probability, which we elsewhere neglected, that the spins end up in a solution state following a dissipation event. The inequality becomes exact in the limit of large $N$ and low dissipation $\rr \ll 1$.  To estimate the reduction in amplification due to dissipation in this limit, we assume a phase step sufficiently narrow that $\avg{\zbar}\approx 1$.  Expanding Eq.~\ref{eq:G0avg_dis} to lowest order in $\rr$ then yields
\begin{equation}
\avg{G_0} \approx 9\left(1 - 4\rr/3\right).
\end{equation}

\section{Experimental implementations}\label{app:implementations}

\subsection{Central spin model with Rydberg atoms}\label{sec:centralSpinModel}
\begin{figure}
    \centering
    \includegraphics[width=\columnwidth]{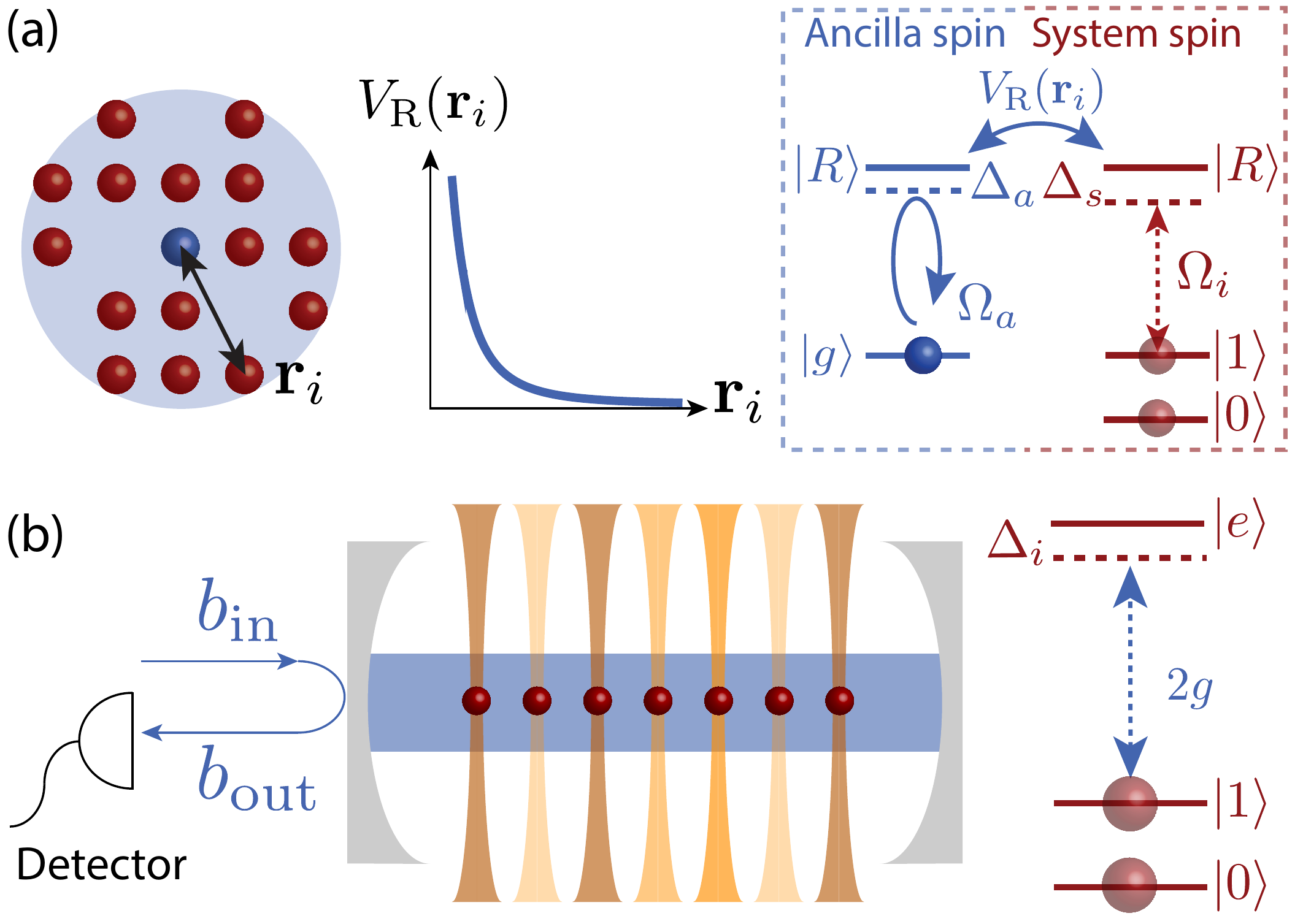}
    \caption{(a) Central spin model realized by Rydberg-dressed atoms (red) interacting with ancilla qubit encoded on a ground-to-Rydberg transition (blue).  
    (b) Central boson model realized by driving one-sided cavity coupled to system spins and heralding on photodetection. 
    }
    \label{fig:experimentalRealizations}
\end{figure}

As a central spin system for encoding subset sum problems, we consider an array of atoms that can be optically coupled to Rydberg states to controllably turn on the interaction Hamiltonian $H_q$ (Eq.~\ref{eq:central_spin}).  The implementation is illustrated in Fig.~\ref{fig:experimentalRealizations}(a). The spins of the system atoms are encoded in two ground states $\ket{0}, \ket{1}$.  The ancilla qubit is encoded using a ground state $\ket{g}$ and a Rydberg state $\ket{R}$, in terms of which we define the spin raising operator $I_+ = \ket{R}\bra{g}$ and lowering operator $I_- = \ket{g}\bra{R}$.  The system is initialized with the ancilla in state $\ket{g}$ and the system spins in state $\ket{\psi_0}$.

To turn on the system-ancilla interactions, the system atoms are individually addressed by control fields that off-resonantly couple state $\ket{1}$ of the $i^\mathrm{th}$ atom to the Rydberg state $\ket{R}$ with Rabi frequency $\Omega_i$ and detuning $\abs{\Delta_s} \gg \Omega_i$.  In this regime, the lowest-order effect of the light on the atomic states is an ac Stark shift given by $\Omega_i^2/(4 \Delta_s)$. Thus we can write the interaction Hamiltonian as
\begin{equation}\label{eq:HR}
H_{R} = \ketbra{R}{R}\sum_i J_i \ketbra{1}{1}_i,
\end{equation}
where
\begin{equation}
J_i = \frac{\Omega_i^2}{4}\left(\frac{1}{\Delta_s - V_R(\vec{r}_i)} - \frac{1}{\Delta_s}\right)
\end{equation}
and $V_R(\vec{r}_i)$ is the Rydberg pair potential between the $i^\mathrm{th}$ system atom and the ancilla.  We choose $V_R$ and $\Delta_s$ to have opposite signs.  If the ancilla is in the Rydberg state, the interaction energy $V_R$ then increases the detuning $\abs{\Delta_s-V_R}$, thereby suppressing the ac Stark shift of atom $i$ by an amount $J_i$.  The result (Eq.~\ref{eq:HR}) is equivalent to the central spin model in Eq.~\ref{eq:central_spin} up to overall energy shifts, with weights $w_i = J_i/J_\mathrm{max}$, where $J_\mathrm{max}$ is the largest of the system-ancilla couplings $J_i$.

The oracle is implemented by simultaneously turning on the couplings $J_i$ and attempting to drive a $2\pi$ pulse on the $\ket{g}\rightarrow\ket{R}$ transition of the ancilla.   The ancilla is driven with a field of Rabi frequency $\Omega_a(t)$, with the pulse shape chosen to ensure that the qubit ends up in its ground state irrespective of whether the pulse is resonant.  This condition is satisfied for a pulse shape~\cite{rosen1932double}
\begin{equation}\label{eq:sech}
\Omega_a(t) = \frac{2\pi}{\tau}\mathrm{sech}\left(\frac{\pi t}{\tau}\right)
\end{equation}
where $\tau$ sets the width of the oracle phase step.  In practice, we must restrict the pulse to a finite window $-t_p/2 < t < t_p/2$, where a duration $t_p \gtrsim 3\tau$ suffices to provide a smooth turn on.  The detuning $\Delta_a$ of the ancilla's control field sets the target weight $\Wtarget = \Delta_a/J_\mathrm{max}$ in the subset sum problem (Eq.~\ref{eq:ss_cond}): for configurations of the system spins with weight $\W = \Wtarget$ in state $\ket{1}$, the ancilla undergoes a $2\pi$ rotation that imparts a geometric phase of $\pi$.

More generally, this protocol produces a unitary transformation
\begin{equation}
U_R = \mathcal{T}e^{-i \int_{-t_p/2}^{t_p/2} H(t)\, dt},
\end{equation}
where we set $\hbar=1$, $\mathcal{T}$ denotes time ordering, and
\begin{equation}
H(t) = H_R + \Omega_a(t)I_x,
\end{equation}
where $I_x = (I_+ + I_-)/2$.
For the hyperbolic secant pulse in Eq.~\ref{eq:sech}, we obtain a $\W$-dependent phase shift $U_R = e^{i\Phi_\th}$ where
\begin{equation}
\Phi_{\th} = 2\arctan\left[2(\Wtarget - \W)/\th\right] + \pi,
\end{equation}
and the width of the phase step is given by $\th = 2\pi / (J_\mathrm{max} \tau)$~\cite{robiscoe1978extension}.

Two effects that can limit the performance of the Rydberg implementation are the finite lifetime $1/\Gamma_R$ of the Rydberg state and residual interactions among the system spins.  The residual interactions between the system spins are smaller than the system-ancilla couplings by a factor of order $(\Omega_i/\Delta_s)^2$ assuming $\abs{V_R(\vec{r}_i)} \gtrsim \abs{\Delta_s}$.  If necessary, these interactions can furthermore be cancelled by an echo procedure in which the control fields $\Omega_i$ are applied again with the signs of $\Delta_s$ and $V_R$ reversed, the latter by tuning the electric field near a F\"{o}rster resonance~\cite{vogt06dipole}.  We therefore neglect residual interactions in our analysis and focus on the limits set by Rydberg decay.

To estimate the requirements for implementing Grover's algorithm while keeping the probability of Rydberg decay small, we define the maximum $\Omega_\mathrm{max}$ of the Rabi frequencies $\Omega_i$ and the dressing amplitude $\epsilon = \Omega_\mathrm{max}/(2\abs{\Delta_s})$.  Our perturbative analysis of the dressing assumes that $\epsilon^2 < 1/n$, where $n$ is the number of system spins.  Let us furthermore assume that the most strongly weighted atom is sufficiently close to the ancilla that $\abs{V_R} \gtrsim \abs{\Delta_s}$, such that its coupling is
\begin{equation}\label{eq:Jmax_ryd}
J_\mathrm{max}\approx \Omega_\mathrm{max}^2/(4\Delta_s)=\epsilon \Omega_\mathrm{max}/2.
\end{equation}
During the oracle pulse, the probability of decay for a system  atom due to the coupling to the Rydberg state will be $t_p \epsilon^2 \Gamma_R$. The worst-case decay probability of the system spins when each spin is in state $\ket{1}$ is $3 \pi n \epsilon^2/\rho \th$, based on the pulse time $t_p = 3 \pi/\th J_\mathrm{max}$. In addition, the error rate due to ancilla decay during the generalized oracle is approximately $\Gamma_R / \Jmax \th$. In the weak dressing limit $n\epsilon^2 \ll 1$, the decay due to the ancilla dominates over the decay of the dressed system spins.

We now present concrete experimental parameters for implementing the central spin model with cesium atoms. Coupling to high-lying Rydberg states is beneficial as the lifetime scales as the cube of the principal quantum number.
By coupling to the $\ket{80P_{3/2}}$ state, we can achieve $\Omega_\mathrm{max} \approx 2\pi \times 10~$MHz with realistic laser parameters~\cite{hankin2014twoatom, borish2020transverse}. The Rydberg interaction strength is given by $V_R(r)=-C_6/r^6 $, where $C_6 \approx 2 \pi \times 7000~\text{GHz}~\upmu \text{m}^6$ for $\ket{80P_{3/2}}$~\cite{sibalic2017ARC}. For a typical distance between neighboring atoms in an optical tweezer array $r_0 \approx 4~\upmu \text{m}$, the interaction shift will be $V_R(r_0) \approx 2 \pi \times 1.7~$GHz. 

The achievable interaction strength in the Rydberg implementation will depend on system size $n$, as the weak dressing condition $\epsilon^2 < 1/n$ puts an upper limit on $J_\mathrm{max}< \Omega_\mathrm{max}/(2 \sqrt{n})$. To give a particular example, for a system size $n=6$, with $n \epsilon^2 = 0.1$ and $\Omega_\mathrm{max}=2\pi\times 10~\text{MHz}$, the interaction strength is $J_\mathrm{max} \approx 650~\text{kHz}$ for $\Delta_s\approx 2\pi\times39~\text{MHz}$. The interaction shift $\abs{V_R(r_0)}>\abs{\Delta_s}$  is large enough to extinguish the light shift of the most strongly coupled atom as we assumed in the preceding analysis. For the state $\ket{80P_{3/2}}$ in cesium, $\Gamma_R \approx 2 \pi \times 0.5~\text{kHz}$, giving the interaction-to-decay ratio $\rho \approx 1200$.

\subsection{Central boson model with atoms in a cavity} \label{sec:centralBosonModel}


As a central boson system for encoding subset sum problems, we consider $n$ spins that are coupled to a cavity of linewidth $\kappa$. We require a dispersive atom-light interaction described by a Hamiltonian
\begin{equation} \label{eq:AtomLightInteraction}
H = c^\dagger c \sum_i J_i \ketbra{1}{1}_i.
\end{equation}
Here, $J_i = g_i^2/\Delta_i$ is the shift of the cavity resonance when the $i^\mathrm{th}$ spin is flipped, in terms of the vacuum Rabi frequency $g_i$ and the detuning $\Delta_i \gg \Gamma_e$ of the cavity from resonance with a transition $\ket{1}\rightarrow\ket{e}$ of linewidth $\Gamma_e$ [Fig.~\ref{fig:experimentalRealizations}(b)]. 

To implement the oracle, the cavity is driven by a weak, narrow-band coherent field $\ket{\alpha}$ of frequency $\omega = \omega_c + \delta$, where $\omega_c$ is the resonance frequency of the bare cavity.  The output and input modes
\begin{equation}
b_\mathrm{out} = \chi b_\mathrm{in}
\end{equation}
are related by the cavity response function~\cite{gardiner1985input}
\begin{equation}\label{eq:chi}
\chi = -\frac{\kappa/2 + i \delta'}{\kappa/2 - i\delta'},
\end{equation}
where $\delta' = \delta - J_{\text{max}} \W$, assuming that cavity losses are negligible compared with transmission. The weighted sum $\W$ is defined as in App.~\ref{sec:subsetsum} using weights determined by the couplings of each spin to the cavity, $w_i = J_i/\Jmax$. The choice of detuning of the drive field from bare cavity resonance $\delta$ sets the target weight $\Wtarget = \delta/\Jmax$ for the subset sum problem. This can be tuned to specifically implement the partition problem (see App.~\ref{sec:subsetsum}).

More generally, we can also account for a photon loss rate $\Gamma_a$, including any absorption by the atoms, by letting
\begin{equation}
\delta' = \delta - \Jmax \W + i \Gamma_a/2.
\end{equation}

The magnitude and phase of the cavity response function $\chi$ determine, respectively, the probability $\abs{\chi}^2$ of successfully detecting the ancilla photon and the resulting oracle phase shift.  On resonance, the magnitude of the response function is
\begin{equation}\label{eq:abs_chi}
\abs{\chi(0)} = \frac{\kappa-\Gamma_a}{\kappa + \Gamma_a},
\end{equation}
which yields a detection probability $\abs{\chi}^2\approx 1- 4 \Gamma_a/\kappa$ for small $\Gamma_a/\kappa$.  The phase shift is given by
\begin{equation}
\Phi(\W) \equiv \arg\left[\chi\right] = 2\arctan(2\delta'/\kappa) + \pi.
\end{equation}
The phase $\Phi$ increases from $0$ to $2\pi$ in a step of characteristic width $\kappa$, assuming low losses $\Gamma_a \lesssim \kappa/2$, as a function of the atom-dependent detuning between the drive and cavity resonance.  We parameterize the step width by the dimensionless value $\gamma = \kappa/\Jmax$.

To apply the oracle $U_\gamma$, we initialize the system in a product state of the atoms, the vacuum field in the cavity, and a weak, narrow-band coherent state in the input mode: 
\begin{equation}
    \ket{\Psi}=\ket{\psi_0}\ket{0_c}\ket{\alpha_{b_{\text{in}}}}.
\end{equation}
The coherent field leaks through the input mirror into the cavity mode, where the light and atoms interact according to Eq.~\ref{eq:AtomLightInteraction}, then leaks into the output mode $b_{\text{out}}$. After a time $t\gg 1/ (\Delta\omega)\gg1/\kappa$, where $\Delta \omega$ is the bandwidth of the input field, the state evolves to
\begin{equation}
    \ket{\Psi_t} = e^{i\alpha \chi b^\dagger_{\text{out}}}\ket{\psi_0}\ket{0_c}\ket{0_{b_{\text{out}}}}
\end{equation}
The action of $e^{i\alpha \chi b^\dagger_{\text{out}}}$ displaces the vacuum state of the output mode \ket{0_{b_{\text{out}}}} such that the detection of a single photon in the output mode heralds the state
\begin{equation}
\langle 1_{b_\mathrm{out}}\ket{\Psi_t} = e^{i\Phi(\W)}\ket{\psi_0}\ket{0_c},
\end{equation}
thus applying the oracle.

As an alternative to the coherent drive and heralding, an ancilla atom can be used as an intracavity single-photon source.  By coupling the ancilla to the cavity via a two-photon transition, with the first leg being a classical field, the cavity can be controllably excited from the vacuum to the single-photon state.  The bosonic mode is thus reduced to two levels $\ket{0}_c, \ket{1}_c$ that are coupled by the control field on the ancilla, so that we effectively recover a central spin model.  The implementation of the oracle then proceeds much as in App.~\ref{sec:centralSpinModel}, by driving a shaped $2\pi$ pulse that returns the ancilla atom to its initial state and the cavity to the vacuum state.  The width $\tau$ of this pulse now controls the step width $\gamma = 2\pi/(\Jmax \tau)$, subject to the requirement that the pulse be short compared to the cavity lifetime.

We now proceed to estimate the cavity parameters required to observe Grover amplification [as in Eq.~(\ref{eq:Gvsmu})], as well as the attainable interaction-to-decay ratio.  Amplifying the probability of solution states requires a phase step narrower than the initial probability distribution $P(\W)$, which in turn requires strong atom-light coupling.  In particular, we will show that the single-atom cooperativity $\eta = 4g^2/(\kappa\Gamma_e)$ sets an upper bound on the dispersive cavity shift $\Jmax$ achievable at low photon loss rate $\Gamma_a < \kappa$, and hence a lower bound on the dimensionless step width $\gamma = \kappa/\Jmax$ in the driven cavity.

The lower bound on the step width $\gamma$ arises because increasing the dispersive coupling $\Jmax$ comes at the cost of increased chance of atomic absorption.  In the worst-case scenario where all $n$ atoms are in state $\ket{1}$ in the scheme of Fig.~\ref{fig:experimentalRealizations}(b), atomic absorption produces a photon loss rate
\begin{equation}\label{eq:GammaA}
\Gamma_a = \Gamma_e\sum_{i=1}^n \frac{g_i^2}{\Delta_i^2} = \Gamma_e \Jmax^2\sum_{i=1}^n \frac{w_i^2}{g_i^2}
\end{equation}
in terms of the weights $w_i$.  While each weight can be tuned via either the atom-cavity coupling $g_i$ or the detuning $\Delta_i$, the latter is preferable because it allows all atoms to benefit from the maximum cavity cooperativity.  Thus we set $g_i\equiv g$ to be maximal for all atoms, reducing Eq.~\ref{eq:GammaA} to
\begin{equation}\label{eq:abs_over_kappa}
\frac{\Gamma_a}{\kappa} = \frac{\Gamma_e\kappa}{\gamma^2g^2}\sum_{i=1}^n w_i^2 = \frac{4 n\wrms^2}{\eta\gamma^2},
\end{equation}
where $\wrms^2$ represents the mean-squared value of weights and is given by $\wrms^2=1/3$ for weights drawn from a uniform distribution $w_i \in (0,1]$.  Thus, keeping photon loss small ($\Gamma_a/\kappa \lesssim 1$) requires a step width $\gamma \gtrsim \sqrt{n/\eta}$.

Equation~(\ref{eq:abs_over_kappa}) gives the decay parameter $\rr = \Gamma_a/\kappa$ necessary to determine the single-cycle amplification in Eq.~\ref{eq:G0avg_dis}.  Notably, we can re-express the decay parameter in terms of the variance $\sigma^2 = n\wrms^2/\gamma^2$ of the normalized weighted spin $\mu = 2(\Wtarget-\W)/\gamma$ and the cooperativity:
\begin{equation}
\rr = \frac{4\sigma^2}{\eta}.
\end{equation}
Achieving amplification requires $\sigma^2 > 1$, i.e., the probability distribution of $\W$ should be broader than the width $\gamma$ of the phase step. To achieve this condition at low loss $\rr<1$, we require strong coupling $\eta \gg 1$.  This requirement is corroborated by plots of the amplification versus step width for various cooperativities in Fig.~\ref{fig:decay}. The maximum achievable single-cycle amplification, shown in Fig.~\ref{fig:decay}(c), becomes larger than 1 for $\eta \gtrsim 50$.  This condition can be satisfied in state-of-the-art optical cavities, where the highest cooperativities achieved are $\eta \sim 10^2$~\cite{colombe2007strong,wolke2012cavity}, at scalable system size $n$.
 
Achieving substantial quantum speedups requires operating in the ultrastrong coupling regime $\eta \gg n$ to reach step widths $\gamma \ll 1$. A cooperativity as high as $\eta = 4\times 10^8$ has been achieved by coupling circular Rydberg atoms to a superconducting millimeter-wave cavity~\cite{haroche2006exploring}, with $(g,\kappa,\Gamma) = 2\pi\times(2.5\times 10^4,1.4,4.4)~\mathrm{Hz}$.  To access this high cooperativity, both spin states $\ket{0},\ket{1}$ must be Rydberg states with finite lifetime $\Gamma^{-1}$, and the dominant decay channel is then atomic decay rather than photon loss, resulting in an interaction-to-decay ratio $\rho \approx \Jmax/(n\Gamma)$. The detunings $\Delta_i$ should be set to maximize the couplings, up to $\Jmax = \epsilon g$, where $\epsilon \equiv g/\mathrm{min}(\Delta_i)$ is limited by the requirement $n\epsilon^2 < 1$ to avoid absorption of the photon.  Fixing $n\epsilon^2 = 0.1$ allows an interaction-to-decay ratio $\rho \approx 2\times 10^3/n^{3/2}$ for the parameters of Ref.~\cite{haroche2006exploring}.  

\bibliography{references}

\end{document}